\documentclass[twocolumn,amsmath,amssymb,nofootinbib,superscriptaddress]{revtex4-2}
\usepackage{graphics,amssymb,amsmath,epsfig,color,textgreek}
\usepackage{graphicx}
\usepackage[dvipsnames]{xcolor}
\usepackage{dcolumn}
\usepackage{bm}
\usepackage[colorlinks=true,citecolor=cyan]{hyperref}
\hypersetup{colorlinks=true,citecolor=cyan,linkcolor=red,urlcolor=magenta}
\usepackage{braket}
\usepackage[normalem]{ulem}
\usepackage{cancel}
\usepackage{xfrac} 
\usepackage{amsfonts}
\usepackage{amsmath}
\usepackage{amssymb}
\usepackage{physics}
\usepackage{svg}

\newtheorem{theorem}{Theorem}
\newtheorem{definition}{Definition}
\newtheorem{lemma}{Lemma}

\newcommand{\kgroup}{\mathcal{K}}
\newcommand{\galg}{\mathfrak{g}}
\newcommand{\kalg}{\mathfrak{k}}
\newcommand{\malg}{\mathfrak{m}}
\newcommand{\halg}{\mathfrak{h}}
\newcommand{\ham}{\mathcal{H}}
\newcommand{\LK}[1]{\textcolor{black}{#1}}
\newcommand{\EK}[1]{\textcolor{black}{#1}}
\newcommand{\EKK}[1]{\textcolor{black}{#1}}

\newcommand{\bigO}{\mathcal{O}}
\newcommand{\redtext}[1]{\textcolor{red}{#1}}

\newcommand{\RTwo}[1]{\textcolor{black}{#1}}

\newcommand{\RThree}[1]{\textcolor{black}{#1}}

\begin{document}


\title{Fixed Depth Hamiltonian Simulation via Cartan Decomposition}

\author{Efekan K\"okc\"u}
\email{ekokcu@ncsu.edu}
\affiliation{Department of Physics, North Carolina State University, Raleigh, North Carolina 27695, USA}

\author{Thomas Steckmann}
\affiliation{Department of Physics, North Carolina State University, Raleigh, North Carolina 27695, USA}

\author{Yan Wang}
\affiliation{Computational Sciences and Engineering Division,Oak Ridge National Laboratory, Oak Ridge, TN 37831, USA}

\author{J.~K.~Freericks}
\affiliation{Department of Physics, Georgetown University, 37th and O Sts. NW, Washington, DC 20057 USA}

\author{Eugene F. Dumitrescu}
\email{dumitrescuef@ornl.gov}
\affiliation{Computational Sciences and Engineering Division,Oak Ridge National Laboratory, Oak Ridge, TN 37831, USA}

\author{Alexander F. Kemper}
\email{akemper@ncsu.edu}
\affiliation{Department of Physics, North Carolina State University, Raleigh, North Carolina 27695, USA}

\date{\today}

\begin{abstract}
 \LK{Simulating quantum dynamics on classical computers is challenging for large systems due to the significant memory requirements. Simulation on quantum computers is a promising alternative, but fully optimizing quantum circuits to minimize limited quantum resources remains an open problem. We tackle this problem presenting a constructive algorithm, based on Cartan decomposition of the Lie algebra generated by the Hamiltonian, that generates quantum circuits with time-independent depth. We highlight our algorithm for special classes of models, including Anderson localization in one dimensional transverse field XY model, where a $\mathcal{O}(n^2)$-gate circuits naturally emerge. Compared to product formulas with significantly larger gate counts, our algorithm drastically improves simulation precision. In addition to providing exact circuits for a broad set of spin and fermionic models, our algorithm provides broad analytic and numerical insight into optimal Hamiltonian simulations.}
\end{abstract}
\maketitle

Constructing arbitrary unitary operations as a sequence of one and two-qubit gates is the task of {\em unitary synthesis} which has applications from quantum state preparation (e.g. via the unitary coupled cluster formalism \cite{cooper2010benchmark,  lee2018generalized}) to quantum arithmetic logic. A paradigmatic problem  \cite{bauer2020quantum,bassman2021simulating} is the unitary synthesis of time evolution under a time-independent Hamiltonian $\mathcal{H}$. 
Hamiltonian evolution plays a key role in simulating quantum systems on quantum computers \cite{feynman1982simulating,Lloyd1996,abrams1997simulations,zalka1998simulating,jordan2012quantum} and thus has spurred recent interest in order to solve difficult problems beyond the scope of classical computing.
It involves solving $i \frac{d}{dt}\ket{\psi(t)} = \ham \ket{\psi(t)} $via the unitary $U(t)=e^{-i\mathcal{H}t}$, which yields $\ket{\psi(t)} = U(t)\ket{\psi(t=0)}$. While the circuit complexity for an arbitrary unitary grows exponentially with the number of qubits, there are efficient product formulas \cite{Lloyd1996,Haah2021,Childs2021}, series expansions \cite{Berry2015}, and other techniques \cite{Low2017, Low2018, kalev2021quantum} for Hamiltonian simulation. 

Despite these algorithms'  efficient asymptotic performance, the fast fidelity decay with respect to circuit depth before error correction prevents useful Hamiltonian simulation in near term hardware \cite{preskill2018quantum}. Reducing the circuit depth required for simulations remains of interest and recent works have begun to incorporate additional problem information such as algebraic properties, system symmetries \cite{Tran2020}, and initial state properties \cite{Sahinoglu2020} to further improve Hamiltonian time-evolution. Orthogonally, variational approaches have been used to approximate the time evolution \cite{cirstoiu2020variational}, but the approximation worsens with increasing time.

Concurrent with the above synthesis techniques, Cartan decomposition emerged \RThree{as a useful tool in the areas of quantum control} \cite{RefB7,d2007introduction} and time evolution \cite{RefB6}. An optimal unitary synthesis of arbitrary two-qubit operations based on the Cartan decomposition has emerged as the state-of-the-art technique \cite{vidal2004universal}. For larger unitaries, Refs.~\onlinecite{khaneja2005constructive,quantum_symmetries, drury2008quantum_shannon, dagli}, \RThree{have formally laid out} how any element in $SU(2^n)$ can be 
decomposed, although these methods generically require exponential circuit depth for \LK{{\em arbitrary}} unitaries
\RThree{and recursive algorithms\cite{khaneja2005constructive,drury2008quantum_shannon, dagli}}. The product factorization works as follows: consider a \RThree{generic}
time-independent Hamiltonian for $n$ qubits (or $n$ spin-\sfrac{1}{2} particles)
\begin{align}
    \ham = \sum_j H_j \sigma^j,
\label{eq:hamiltonian}
\end{align}
where $H_j$ are \RThree{real} coefficients and $\sigma^j$ are Pauli string operators: \textit{i.e.}, elements of the $n$-site Pauli group $\mathcal{P}_n=\{I,X,Y,Z\}^{\otimes n}$.  Ref.~\onlinecite{khaneja2005constructive} recursively obtains a factorization of the time-evolution unitary as
\begin{align}\label{eq:U(t)product}
    U(t) = e^{-i\mathcal{H}t}=\prod_{\bar{\sigma}^i \in \mathfrak{su}(2^n)} e^{i \kappa_i \bar{\sigma}^i},
\end{align}
with, in the general case, $\bigO(4^n)$  angles $\kappa_i$ for the Pauli strings $\bar\sigma^i$ that form a basis for the Lie algebra $\mathfrak{su}(2^n)$. 

We now provide a constructive decomposition algorithm for Hamiltonian simulation with depth independent of simulation time.
\RTwo{We begin by applying Cartan decomposition on a subalgebra of $\mathfrak{su}(2^n)$ generated from the Hamiltonian.}\RTwo{We further simplify the subsequent problem of finding the parameters $\kappa_i$ to locating a \emph{local} \RTwo{extremum}, rather than global minimum, of a cost function by extending the method given in }Refs.~\cite{helgason2001differential, earp2005constructive}. \RTwo{This extension allows us to directly generate a circuit,}
calculate the cost function and its gradient. The algorithm is applicable to \emph{any} model without limitations of locality, although the scaling varies depending on the model, \RTwo{and we provide software to do
so\cite{cartan_code}.}

For certain classes of models (termed ``fast-forwardable\cite{gu2021fastforwarding}'') such as spin models which can be mapped to non-interacting fermion models \cite{chapman2020characterization}\RTwo{,} the circuit complexity and calculation of the cost function scales polynomially in the system size.
%
To illustrate our algorithm, we use it to 
time evolve a 10-site random transverse field XY (TFXY) model and compare the result
to a Trotter approach to illustrate the dramatic improvements obtained. 

\begin{figure}[b]
    \includegraphics[clip=true,trim=40 0 30 50,width = 0.95\columnwidth]{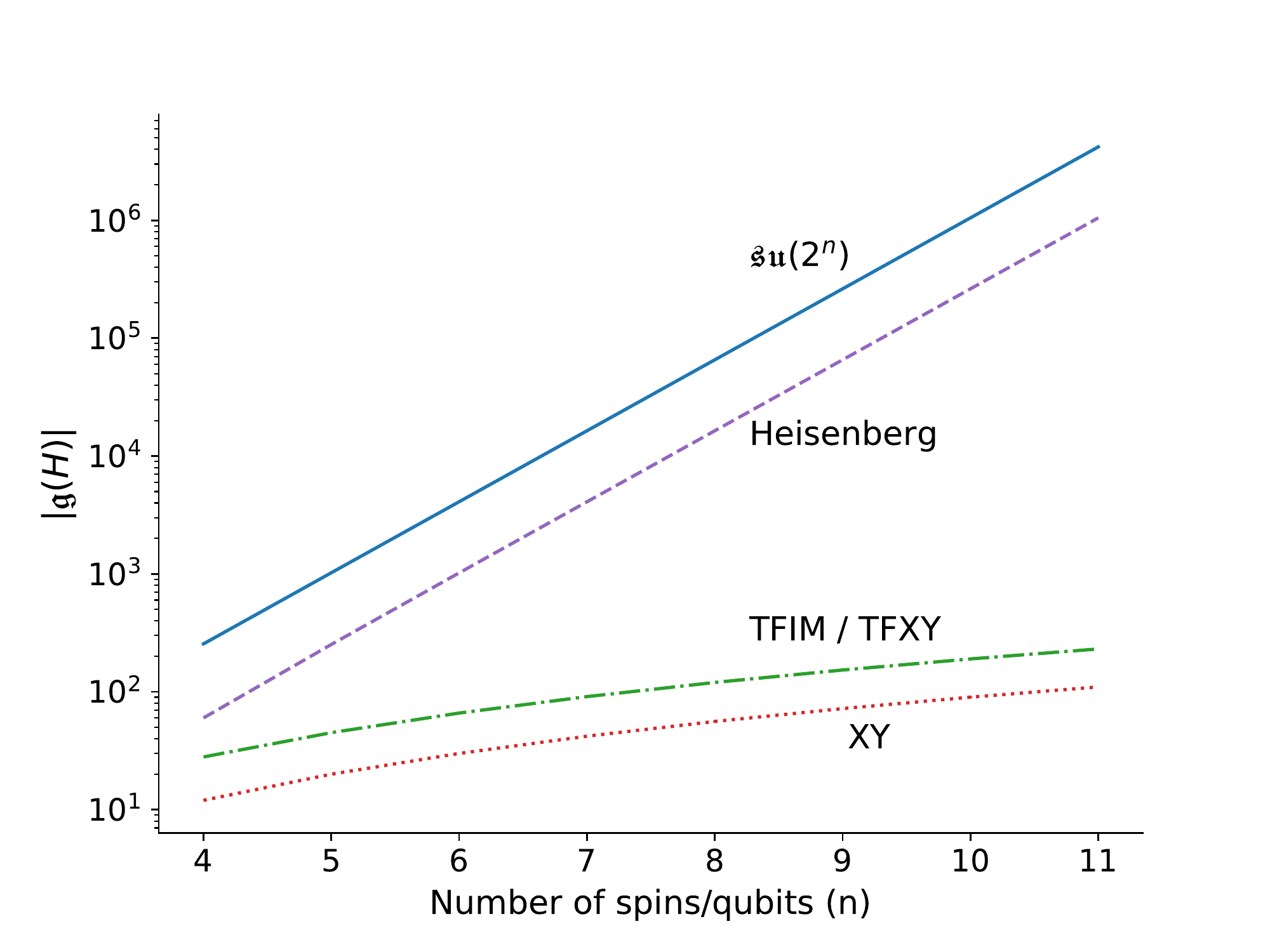}
    \caption{Hamiltonian algebra dimensions of the nearest-neighbor Heisenberg, XY, TFIM and TFXY models, and dimension of full $\mathfrak{su}(2^n)$ for comparison to the generic case. The dimensions can exactly be calculated as $|\galg(\mathrm{Heisenberg})| = 4^{n-1}-4$, $|\galg($TFIM$)| = |\galg($TFXY$)| = n(2n-1)$ and $|\galg($XY$)| = n(n-1)$. }
    \label{fig:scaling}
\end{figure}
{\em Hamiltonian Algebra}---For a given Hamiltonian, we determine whether the entirety of $\mathfrak{su}(2^n)$ is necessary for the expansion in Eq.~\eqref{eq:U(t)product}, or whether a subset suffices. 
The Baker-Campbell-Hausdorff theorem states that only nested commutators of the individual terms in the Hamiltonian appear in the final exponent.  This leads us to the \textbf{first step of our algorithm:} Using the expansion of the Hamiltonian in terms of the Pauli terms $\sigma^j$ in Eq.~(\ref{eq:hamiltonian}), find the closure (under commutation) of the set of those Pauli terms. This closure forms a basis for the {\em Hamiltonian algebra}, which we denote as $\galg(\ham)$, and which is a subalgebra of $\mathfrak{su}(2^n)$ \cite{[{Within the context of control theory, the Hamiltonian algebra is also referred to as the dynamical algebra: }] d2007introduction}.
We can now restrict the expansion in Eq.~\eqref{eq:U(t)product} to only the elements of $\galg(\ham)$. 

\EK{We now provide some examples on the scope and limitations of our resource cost across selected spin Hamiltonians.} Fig.~\ref{fig:scaling} illustrates the dimension of the Hamiltonian algebra $|\galg(\ham)|$ for various models of interest as a function of system size $n$, where $|\cdot|$ denotes the dimension of the algebra.  The dimensions of the Hamiltonian algebra for the $n$-site nearest-neighbor \RTwo{XY}, transverse field Ising (TFIM) \LK{and TFXY} models are $|\galg(\mathrm{XY})|=n(n-1)$ and $|\galg(\mathrm{TFIM})|=|\galg(\mathrm{TFXY})|=n(2n-1)$; these scale \textit{quadratically} with the number of qubits $n$. On the other hand, $|\galg(\ham)|$ for the nearest-neighbor Heisenberg model scales exponentially, $|\galg(\mathrm{Heisenberg})|=4^{n-1}-4$. We observe a similar exponential growth in TFXY and TFIM models with longer range interactions. However, even in these cases, $|\galg(\ham)|$ is \EK{a constant factor} smaller than $|\mathfrak{su}(2^n)|$, \EK{providing a commensurate improvement in circuit depth over the generic} case studied in Ref.~\cite{khaneja2005constructive}.

{\em Cartan Decomposition} --- 
We must now determine the parameters $\kappa_i$ in the $\galg(\ham)$-restriction of Eq.~\eqref{eq:U(t)product}. The Cartan decomposition and related methods in Ref.~\onlinecite{khaneja2005constructive, helgason2001differential, earp2005constructive} provide the necessary tools to do so. We briefly review the Cartan decomposition and the ``KHK theorem''. 
%
\begin{definition}\label{def:cartan_decomposition}
    \RThree{A} \textbf{Cartan decomposition} of
    \RThree{a Lie algebra}
    $\mathfrak{g}$ is defined as an orthogonal split $\mathfrak{g} = \mathfrak{k} \oplus \mathfrak{m}$ satisfying
\begin{align}
    [\kalg, \kalg] &\subset \kalg, &
    [\malg, \malg] &\subset \kalg, &
    [\kalg, \malg] &= \malg, 
    \label{Cartan Split}
\end{align}
and denoted by $(\galg,\kalg)$. \RThree{A} \textbf{Cartan subalgebra} denoted by $\halg$ refers to a \RThree{maximal Abelian algebra within} $\malg$.
\end{definition}
\begin{figure*}[t]
    \includegraphics[width = 0.95\textwidth]{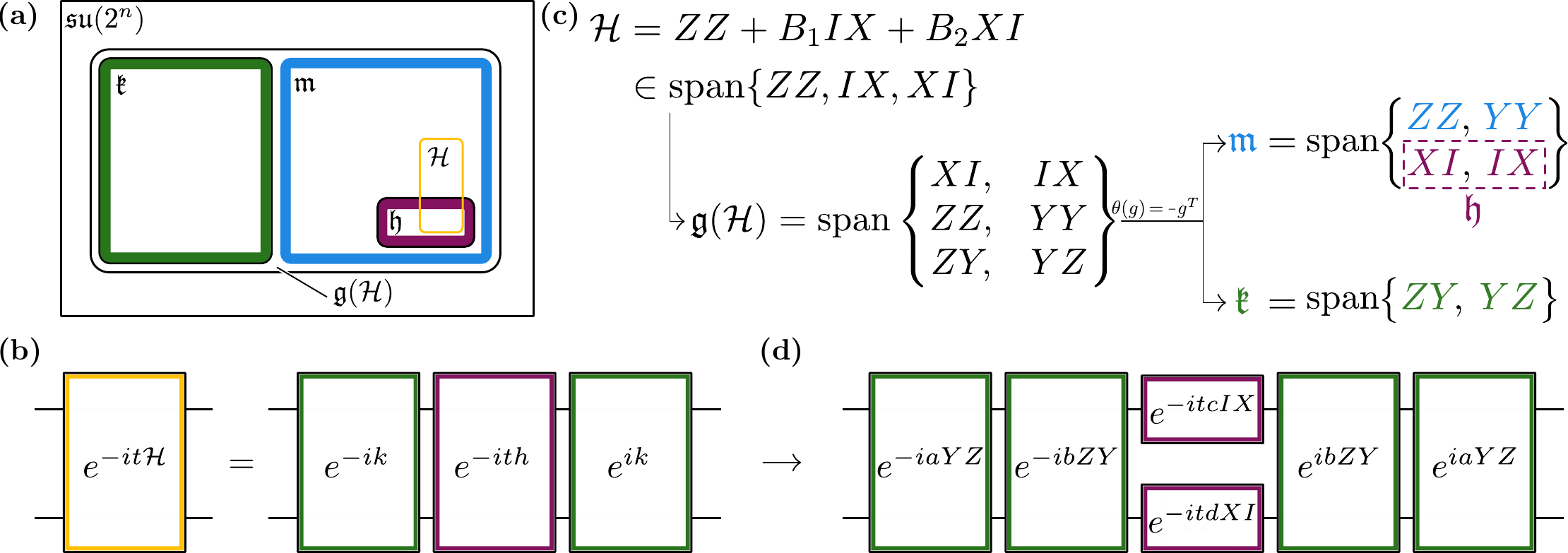}
    \caption{(a) Schematic relationship of the Hamiltonian algebra $\galg(\mathcal{H})$ and its partitioning into a subalgebra $\kalg$, its compliment $\malg$, and the Cartan subalgebra $\halg$.
    (b) KHK decomposition (Theorem~\ref{th:KHK}) applied to a time evolution operator generated by an element of $\malg$.
    (c) Hamiltonian algebra $\galg(\ham)$ for the 2 site TFIM and the Cartan decomposition generated by the involution $\theta(\galg)=-\galg^T$. Here we list
    the bases that span $\galg(\ham)$ and its
    Cartan decomposition.
    (d) Decomposed time evolution for the 2-site TFIM model.}
    \label{fig:schematic}
\end{figure*}
\RTwo{We will replace $\galg$ in Def. \ref{def:cartan_decomposition} above with $\galg(\ham) \subseteq \mathfrak{su}(2^n)$ for a given $n$-spin Hamiltonian. 
} 

In practice \EK{finding a Cartan decomposition by directly using Def. \ref{def:cartan_decomposition} and picking basis elements one by one is difficult. Instead} the Lie subalgebra is partitioned \EK{into $\kalg$ and $\malg$} by an involution: \EK{i.e. a} Lie algebra homomorphism taking $\theta: \galg \to \galg$, which satisfies $\theta(\theta(g)) = g$ for any $g \in \galg$ and preserves all commutators. \EK{Then by using the involution, one can split the algebra by defining sub-spaces via $\theta(\mathfrak{k}) = \mathfrak{k}$ and $\theta(\mathfrak{m}) = -\mathfrak{m}$, which is equivalent to Def. \ref{def:cartan_decomposition}. We discuss further details of involutions in the SI.} 

A consequence of  Cartan decomposition, which we will use to synthesize Hamiltonian evolution unitaries, is \RThree{an extension of the} \LK{``KHK''} theorem:
\begin{theorem}
    Given a Cartan decomposition $\galg = \mathfrak{k} \oplus \mathfrak{m}$ \RThree{and a non-degenerate invariant bilinear form $\langle .,.\rangle$ on $\galg$ then} for any $m \in \malg$ there exists a $K \in e^{i\kalg}$ and an $h \in \halg$, such that
    \begin{align}
        m = K h K^\dagger,
    \end{align}
    \label{th:KHK}
\end{theorem}
%
%
\RThree{where we have generalized the KHK theorem to any Lie algebra that has a non-degenerate invariant bilinear form. 
This statement is proven by construction via Thm.~\ref{thm:earppachos}~\cite{supplementary}. We use $\langle A,B \rangle = \mathrm{tr}(AB)$, which is proportional to the Killing form in $\mathfrak{su}(2^n) \supset \galg(\ham)$, and is therefore guaranteed to be  non-degenerate due to semi-simplicity of $\mathfrak{su}(2^n)$. Moreover, it is invariant and symmetric due to the cyclic property of the trace.}

We can now describe the \textbf{second step of our algorithm:} Find a Cartan decomposition of $\galg(\ham)$ such that $\ham \in \malg$ \LK{(in practice, one finds an involution)}\RTwo{, and find a Cartan subalgebra $\halg \subseteq \malg$}. A direct application of Theorem \ref{th:KHK} with $\ham = K h K^\dagger$ then leads to the desired unitary for time-evolution
\begin{align}\label{eq:time_evolution}
 U(t) = e^{-i \ham t} = Ke^{-ih t}K^\dagger.
\end{align}
\RTwo{Since $\halg$ is Abelian, each Pauli string in $h \in \halg$ commutes, and therefore a quantum circuit for $e^{-ith}$ can easily be constructed. This reduces the circuit construction problem to finding 
$K$, which we address below}.

\EK{As long as an involution is found such that $\theta(\ham) = -\ham$, this method is applicable to any Hamiltonian $\ham$. Specifically, for} the models discussed in Fig.~\ref{fig:scaling}, this step is achieved by using the involution $\theta(g) = -g^T$ \EK{which is an AI type Cartan decomposition for $\mathfrak{su}(2^n)$.}
\EK{This involution works because the listed 
models have time reversal symmetry \cite{quantum_symmetries}.}
\RTwo{We then construct $\halg$ by choosing an element
of $\malg$ randomly (or with certain symmetries if desired) and finding all the elements in $\malg$ that are mutually commutative with the chosen element and each other.} We discuss further details of finding involutions and Cartan subalgebras in \RTwo{\cite{supplementary}}.

Note that the simulation time $t$ in Eq.~\eqref{eq:time_evolution} enters as an independent parameter, and does not alter the structure of $K$ or $h$. 
This means that new parameters do not need to be found for different simulation times
(although this situation may change for time-dependent Hamiltonians).

{\em Determining Parameters} --- 
%
We provide the following theorem to determine the group element $K$ in Eq.~\eqref{eq:time_evolution}, \RTwo{which is an improved version of Lemma 6.3 (iii) in \cite{helgason2001differential} and Eq. \redtext{18} in \cite{earp2005constructive}: }
\begin{theorem}\label{thm:earppachos}
\RTwo{Assume a set of coordinates $\Vec{\theta}$ in a chart of the Lie group $e^{i \kalg}$.}
For $\ham \in \malg$, define the function $f$
\begin{align}\label{eq:Earp_function}
    f(\Vec{\theta}) = \langle K(\Vec{\theta}) v K(\Vec{\theta})^\dagger, \ham \rangle,
\end{align}
where $\langle .,. \rangle$ denotes \RThree{a non-degenerate invariant bilinear form on $\galg$}, and $v \in \halg$ is an element whose exponential map is dense in $e^{i \halg}$. Then for any {\em local extremum} of $f(\Vec{\theta})$ denoted by $\Vec{\theta}_c$, and defining the critical group element $K_c = K(\Vec{\theta}_c)$, we have
\begin{align}
    K(\Vec{\theta}_c)^\dagger \ham K(\Vec{\theta}_c)= K_c^\dagger \ham K_c  \in \halg.
    \label{eq:critical_k}
\end{align}
\end{theorem}

According to the theorem, we only need to find a local extremum of $f(\Vec{\theta})$, without determining the resulting $h \in \halg$. This is achieved by using $v$ such that $e^{itv}$ is dense in $e^{i\halg}$;
this is sufficient to represent the entirety of $\halg$. This reduces the number parameters from  $|\kalg| + |\halg|$ to $|\kalg|$. \RTwo{Since we consider single Pauli strings as basis elements, we can choose $v = \sum_i \gamma_i h_i$ where the $h_i$ are basis elements of $\halg$, and the $\gamma_i$ are mutually  irrational\cite{earp2005constructive}. After determining $K_c$, the $h \in \halg$ in Eq.~\eqref{eq:time_evolution} is then obtained via Eq.~\eqref{eq:critical_k}.} 
Further details and the proof of the theorem are discussed in \RTwo{\cite{supplementary}}.

\RTwo{Because the parametrization does not need to cover the entire $e^{i\kalg}$, }there is a choice in how to represent the group element $K$ \EK{in Thm.~\ref{thm:earppachos}}. While   Refs.~\onlinecite{khaneja2005constructive,earp2005constructive}  use $K = \exp( i \sum_i \alpha_i k_i)$, we express it as a factorized product
\begin{align}
    K = \prod_{i} e^{i a_i k_i},
    \label{eq:kform}
\end{align}
where $k_i$ is an element of the Pauli string basis for $\kalg$. 
\RTwo{The representation Eq.~\eqref{eq:kform} does not always cover $e^{i \kalg}$ fully, except in some specific cases \cite{wei_norman, Izmaylov2020}, but this is not necessary\cite{supplementary}.} 
However, using Eq.~\eqref{eq:kform} has three benefits. First, the gradient of Eq.~\eqref{eq:Earp_function} can be obtained analytically at any point in contrast to the complicated derivative of the exponential map $\exp(i\sum_i \alpha_i k_i)$; second, this allows us to apply $K$ on $v$ and \RTwo{$\ham$} exactly \RTwo{\cite{supplementary}}; and third, since a circuit implementation for exponentiated individual Pauli strings is known \cite{pauli_circ1,pauli_circ2},
we avoid the need for further decomposition of $K$.

\begin{figure*}[t]
    \includegraphics[width = \linewidth]{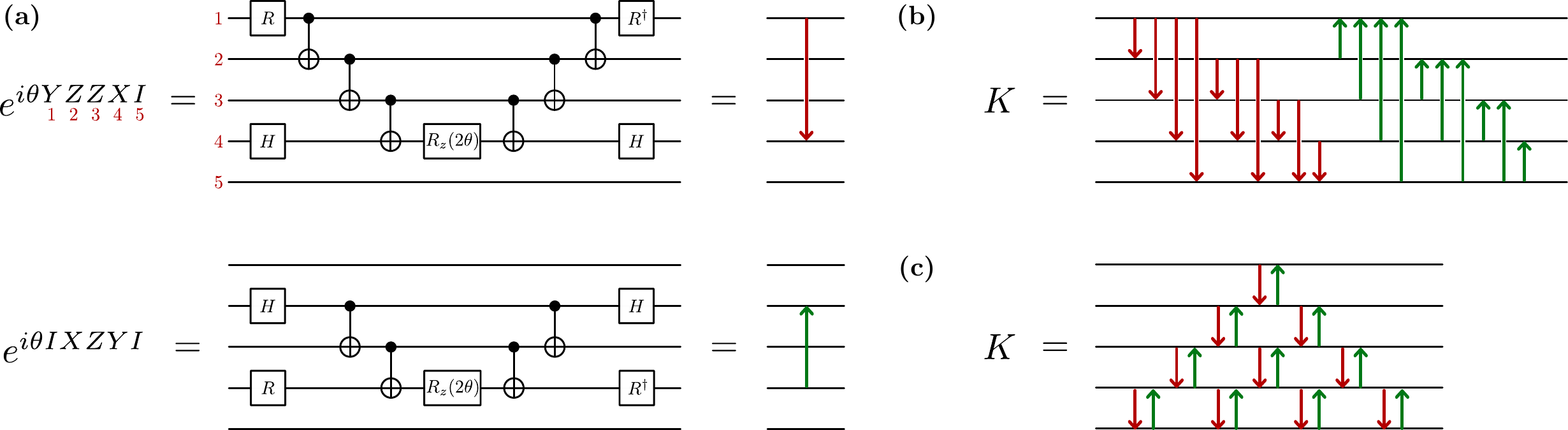}
    \caption{(a) Circuit implementation of the given exponentials of Pauli strings, and the compact arrow notation. The $R$ gate shown here is $R_x(\pi/2)$. (b/c) Unoptimized/optimized circuit for $K$ in an $n=5$ site TFXY model (this system size is chosen for illustrative purposes). The circuits have $\bigO(n^3)$ \EK{(80)}
    and $\bigO(n^2)$ \EK{(20)} CNOT gates, respectively.}
    \label{fig:pauli_circ}
\end{figure*}

We now reach the \textbf{third step of our algorithm:} Minimize Eq.~\eqref{eq:Earp_function} over the parameters $a_i$ in $K$ in Eq. \eqref{eq:kform} to find $K \in e^{i\kalg}$. In this work, we use a standard BFGS optimization routine. 
%
Calculating Eq.~\eqref{eq:Earp_function}, its gradient \RTwo{and obtaining $h \in \halg$ by using} 
Eq.~\eqref{eq:kform} require $\bigO(|\kalg||\malg|)$, $\bigO(|\kalg|^2 |\malg|)$ \RTwo{and $\bigO(|\kalg||\malg|)$} operations, respectively.
For models where $|\galg(\ham)|$ is quadratic
in the number of spins, these become
$\bigO(n^4)$, $\bigO(n^6)$ and $\bigO(n^4)$ \cite{supplementary}.

\EK{In summary, our algorithm can be listed as the following three steps: 
\begin{enumerate}
    \item Construct Hamiltonian algebra $\galg(\ham)$
    \item Find a suitable Cartan decomposition (or involution) such that $\ham \in \malg$, and construct \RTwo{a} Cartan subalgebra $\halg$.
    \item \RTwo{Find a \emph{local} extremum of} $f(\Vec{\theta})$
    by representing $K$ as \RTwo{in Eq.~\eqref{eq:kform}, obtain $h \in \halg$ via Eq.~\eqref{eq:critical_k}},
    and then construct the circuit.
\end{enumerate}}

Fig.~\ref{fig:schematic} is a schematic illustration of the algorithm. Panel (a) shows the relationships
between $\mathfrak{su}(2^n)$, the Hamiltonian $\ham$, the Hamiltonian algebra $\galg(\ham)$, and its Cartan decomposition. Panel (b) shows the resulting factorization of the time-evolution operator. Panels (c) and (d) demonstrate steps one and two of our algorithm for a simple two-site Ising model. In this case, the Hamiltonian terms $\{ZZ,IX,XI\}$ generate a six dimensional Hamiltonian algebra $\galg(\ham)$, which is partitioned into $\kalg$ and $\malg$ via
the involution $\theta$. \EK{Among infinitely many possibilities, there} are two maximal Abelian subalgebras $\halg$ of $\malg$ \EK{that have single Pauli strings as basis elements (rather than a linear combination of them)}, namely $\mathrm{span}\{ZZ,YY\}$ and $\mathrm{span}\{XI,IX\}$; we choose the latter without loss of generality. The resulting factored time-evolution operator is shown in panel (d). 
This factorization is clearly sub-optimal for the Hamiltonian evolution unitary in $SU(4)$, where a minimal 3-CNOT circuit is known \cite{vidal2004universal}; however,
our decomposition algorithm is applicable to any system size.

{\em Application} --- To demonstrate the flexibility of our method, we \LK{simulate 10-site TFXY spin chain with random magnetic field with open boundary conditions:}, with the Hamiltonian 
\begin{align}
    \ham = \sum_{i=1}^{n-1}( X_i X_{i+1} + Y_i Y_{i+1}) + \sum_{i=1}^n b_i Z_i,
\end{align}
where $n=10$ is the number of qubits and the $b_i$ coefficients are \EK{chosen via a normal distribution with zero mean and $\sigma^2$ variance}; we use standard notation for the Pauli spin matrices. We consider a single spin-flip initial state $\ket{\psi} = \ket{\downarrow \uparrow \uparrow \uparrow \uparrow \uparrow \uparrow \uparrow \uparrow \uparrow}$. In the absence of the random magnetic field, this excitation diffuses throughout the system. By increasing the random magnetic field strength, the excitation is prevented from diffusing by \LK{amplitude cancellation due to random phases acquired via probing the random magnetic field, which is called} the Anderson localization
mechanism \cite{andersonloc1}. \EK{Specifically} in one dimension it was shown that any $p$-th moment of the displacement of the excitation has a time independent upper bound $\big< |\hat{N}|^p \big>_t < C$, \EK{ where $C$ is a time independent constant, and} the position operator for the excitation is $\hat{N} = \sum_{r=1}^n (r-1) \frac{1-Z_r}{2}$ \cite{andersonloc2}.

We \LK{first} perform steps one and two of our algorithm. The Cartan decomposition and subalgebra for this model are
\begin{align}\label{eq:cartandec_tfxy}
\begin{split}
     \kalg &= \mathrm{span}\{ \widehat{X_i Y}_{j},\widehat{Y_i X}_{j} \,\big|\, i,j = 1,2,..,n, \: i<j\}, \\
     \malg &= \mathrm{span}\{ Z_j, \widehat{X_i X}_{j},\widehat{Y_i Y}_{j} \,\big|\, i,j = 1,2,..,n, \: i<j\}, \\
     \halg &= \mathrm{span} \{ Z_i \,\big|\, i=1,2,..,n\},
\end{split}
\end{align}
with dimensions $|\kalg| = n(n-1)$,  $|\malg| = n^2$ and $|\halg| = n$, and 
\begin{align}\label{eq:hat_notation}
    \widehat{A_i B_j} = A_i Z_{i+1} Z_{i+2} ... Z_{j-1} B_j.
\end{align}
We then perform step three of our algorithm and find the parameters minimizing Eq.~(\ref{eq:Earp_function}).

Using Eq.~\eqref{eq:kform} generates the circuit shown in Fig.~\ref{fig:pauli_circ}(b), which has $2n(n^2-1)/3$ CNOT gates \LK{(1320 CNOTs for $n=10$)}. As illustrated in
Fig.~\ref{fig:pauli_circ}(c), this circuit
can be further simplified to a circuit with $n(n-1)$ CNOT gates \LK{(180 CNOTs for $n=10$)} \RTwo{\cite{supplementary}}
\LK{We compare the simulation results conducted via our algorithm to Trotter evolutions with varying time steps and fixed depth (fixed number
of CNOTs) that is equal to the optimized ($10$ steps/$180$ CNOTS) and the un-optimized Cartan circuits ($74$ steps/$1332$ CNOTS).}

\newpage

\begin{figure}[t]
    \includegraphics[width = \columnwidth]{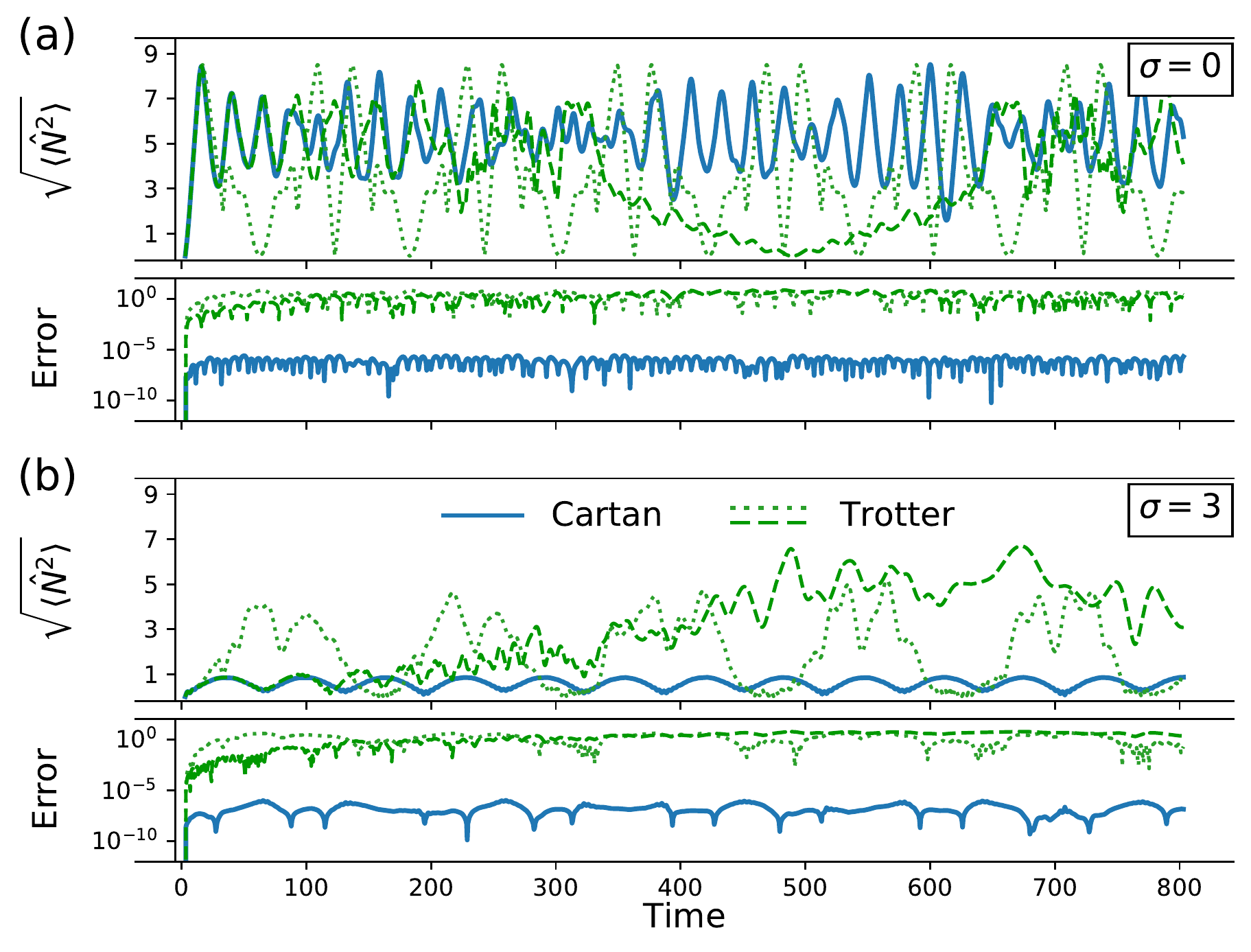}
    \caption{Displacement of the spin excitation $N = \sqrt{\big< \hat{N}^2 \big>}$ and its absolute difference from the exact result $|N-N_{\text{exact}}|$  in the TFXY model with a random Z field, for standard deviation $\sigma=0$ in panel (a) and $\sigma=4$ in panel (b). The excitation becomes trapped around its original position as $\sigma$ increases. The localization is captured to within a small constant error by our Cartan algorithm (solid curves). \LK{The two Trotter decompositions use
    180 (dotted) and 1332 (dashed) CNOTs, which correspond to the CNOT counts of the optimized and non-optimized Cartan circuits, respectively.}}
    \label{fig:anderson_time_evo}
\end{figure}

Fig.~\ref{fig:anderson_time_evo} shows $N = \sqrt{\big< \hat N^2 \big>},$ the RMS position of the single-spin excitation for various values of $\sigma$, as simulated with our algorithm and with Trotter time evolution. We renormalized the Hamiltonian for each standard deviation of the transverse field; \textit{i.e.}, $\ham \rightarrow \frac{\ham}{\sqrt{\tr(\ham^2)}}$ to eliminate any norm dependence of the time evolution.

As expected, the Trotter evolution diverges from the exact result after some time $\tau$, which occurs later if there are more Trotter steps. $\tau$ depends on the standard deviation of disorder $\sigma$ in the magnetic field; the results improve with increasing randomness because
this decreases the relative diffusion probability for the excitation to hop to another site. Nevertheless, for any value of $\sigma$ and
any number of steps, the Trotter evolution eventually diverges from the exact result.


On the other hand, the result from the Cartan decomposition is indistinguishable from the exact solution. We show the error (absolute deviation from the exact result)
for the two methods in Fig.~\ref{fig:anderson_time_evo}.
Except for the earliest times, there are 3-5 orders of magnitude less error for the Cartan decomposition approach compared to 
the Trotter-based approach. The error of the Cartan-based method stems from the non-zero gradient tolerance used in the optimization step of the algorithm (which was chosen to be $10^{-6}$), and does not increase with simulation time, which shows the suitability of this constant-depth circuit for long-time simulations. 
While this particular application is for a free-fermionic model,
the minimal error does not follow from this property. Rather, it stems from the precise factorization
via Cartan decomposition, which is equally applicable
to interacting fermion models. However, a similar tolerance
may lead to larger errors simply due to the larger number
of terms required for the decomposition (Eq.~\ref{eq:kform}). \RTwo{Data for Fig. \ref{fig:anderson_time_evo} is provided at \cite{cartan_data}.}

\RTwo{
{\em Discussion and Applications} --- We have introduced an algorithm based on the Cartan decomposition for synthesizing Hamiltonian time evolution unitaries \RTwo{and provided software to do so
\cite{cartan_code}}. In contrast to previous related approaches \cite{khaneja2005constructive, drury2008quantum_shannon, dagli,earp2005constructive}, the current work develops \textit{explicit} digital quantum circuit constructions for $K$ via a implementable factorized form (Eq.~\ref{eq:kform} and Fig.~\ref{fig:pauli_circ}). An analytic cost function and its derivatives, straightforward circuit construction, and only needing a {\em single} optimization for any time $t$ are several improvements with respect to previous algorithms. We have discussed illustrative and paradigmatic examples where our algorithm's complexity grows polynomially, as in TFIM and TFXY spin-models. Here the polynomial complexity follows a mapping from the spin representation to a non-interacting (free) fermionic representation. In this sense, the Hamiltonian algebra reveals the existence of such a map and is complementary to a recent graph-theoretic approach to identify spin models solvable by fermionization \cite{chapman2020characterization}.
}\RTwo{
This idea has already formed the basis for related compression algorithms\cite{kokcu2021algebraic,*camps2021algebraicb}.
Our work also provides an intuition to understanding heuristic ``variational fast-forwarding'' methods\cite{cirstoiu2020variational,bassman2022constantdepth, berthusen2021quantum}; the scaling of the Hamiltonian algebra indicates an upper bound on the required circuit depth. 
}

\RTwo{
In addition to the applications demonstrated here, we expect our algorithm and its components to find broader use in more quantum computing application areas. First, our method can be applied directly to simulating \emph{both} free and interacting theories and directly deployed on quantum computers. Although the algebra does not scale favorably in the latter case, circuits for interacting fermionic problems can be composed via our technique nevertheless. Given that near term devices scale poorly with circuit depth, and consequently simulation time, employing our algorithm to small systems yields results which we are not aware of other methods achieving \cite{steckmann2021simulating}. Next, a generalization to the unitary coupled cluster (UCC) formalism\cite{cooper2010benchmark, manrique2020momentum, lee2018generalized} is also straightforward. In order to represent the wave function, UCC applies excitations on an ansatz wave function. 
The usual Trotter-based approach to construct circuits to do so does not respect the symmetries inherent in the problem --- this is true for UCC excitations\cite{gard2020efficient}
as well as Hamiltonian evolution\cite{Tran2020} --- this issue can be addressed either
by adding additional symmetry-restoring terms\cite{Tran2020} or
constructing explicit symmetry-preserving circuits\cite{gard2020efficient}.
Since the Cartan decomposition is exact,
it preserves all of the symmetries without further effort,
even though the individual
terms may break symmetries. 
We will detail this application area in a future work. 
This concept of using Cartan decomposition to generate a subcircuit for symmetrized UCC factors --- a portion of a larger problem --- could be applied as a generic quantum compilation routine.}


\RTwo{
Looking forward, we expect that perturbative approaches beginning from either the free or only interacting algebras will enable further progress in the development of Hamiltonian evolution algorithms. Symmetries and other problem structure are naturally expressed in the language of Lie algebras and further developments are required to fully utilize problem structure. Interestingly, preliminary findings indicate that imposing symmetry complicates quantum circuit construction while it reduces the dimension of the Hamiltonian algebra; this interplay between physical symmetry and algebraic analysis for quantum circuits has been recently investigated within the contexts of quantum control theory\cite{RefB1, RefB2, RefB5, RefB8} and symmetry-preserving circuits\cite{economou,tran2021Fastera}, and could be combined with the methods presented here in future work. 
  }


\begin{acknowledgments}
EK, TS, JKF and AFK were supported by the Department of Energy, Office of Basic Energy Sciences, Division of Materials Sciences and Engineering under Grant No. DE-SC0019469. EK
and AFK were also supported 
by the National Science Foundation under Grant No. PHY-1818914.
JKF was
also supported by the McDevitt bequest at Georgetown
University.
EFD acknowledges DOE ASCR funding under the Quantum Computing Application Teams program, FWP number ERKJ347. 
TS was supported in part by the U.S. Department of Energy, Office of Science, Office of Workforce Development for Teachers and Scientists (WDTS) under the Science Undergraduate Laboratory Internship program.
YW acknowledges DOE ASCR funding under the Quantum Application
Teams program, FWP number ERKJ335. 
\end{acknowledgments}

\bibliographystyle{apsrev4-2}
\bibliography{refs.bib}

\clearpage
\onecolumngrid
\appendix

\renewcommand\thefigure{S\arabic{figure}}  
\setcounter{figure}{0}

\section{Hamiltonian Algebras of Certain Models}\label{dimension_appendix}
\subsection{XY Model}

For the 1-D nearest neighbour XY model with open boundary conditions and arbitrary interaction coefficients,  
\begin{align}
    \ham = \sum_{i=1}^{n-1}(a_i X_i X_{i+1} + b_i Y_i Y_{i+1}),
\end{align}
and the Hamiltonian algebra is found to be
\begin{align}
     \galg(\text{XY}) = \mathrm{span}\{ \widehat{X_i X}_{i+a},\widehat{Y_i Y}_{i+a},\widehat{X_i Y}_{i+b},\widehat{Y_i X}_{i+b} \big| \text{$a$ odd, $b$ even, $1 \leq i,i+a,i+b \leq n$}\}.
\end{align}
The dimension of this algebra is calculated as $|\galg(\text{XY})| = 2 \binom{n}{2} = n(n-1)$.

\subsection{TFIM and TFXY Model}

For the 1-D nearest neighbour transverse field XY model with open boundary conditions and free coefficients,
\begin{align}\label{eq:tfxy_ham}
    \ham = \sum_{i=1}^{n-1}(a_i X_i X_{i+1} + b_i Y_i Y_{i+1}) + \sum_{i=1}^{n}c_i Z_i,
\end{align}
the Hamiltonian algebra is found to be
\begin{align}
     \galg(\text{TFXY}) = \mathrm{span} \{ Z_j, \widehat{X_i X}_{j},\widehat{Y_i Y}_{j}, \widehat{X_i Y}_{j},\widehat{Y_i X}_{j} \big|1 \leq i,j \leq n; \: i<j\}.
\end{align}
The same algebra is found for transverse field Ising model, i.e. the $b_i=0$ case for the Hamiltonian given in \eqref{eq:tfxy_ham}. The dimension of this algebra is $|\galg(\text{TFXY})|=n+4\binom{n}{2}=n(2n-1)$.

\subsection{Heisenberg Model}

For the 1-D nearest neighbour Heisenberg model with open boundary conditions and free coefficients,
\begin{align}
    \ham = \sum_{i=1}^{n-1}(a_i X_i X_{i+1} + b_i Y_i Y_{i+1} + c_i Z_i Z_{i+1}),
\end{align}
the Hamiltonian algebra is found to be
\begin{align}
\begin{split}
    \galg(\text{Heisenberg}) = \mathrm{span} \Big ( \{& \text{Pauli strings with } a \text{ many } X,\: b \text{ many } Y,\: c \text{ many } Z \big| a+b,a+c,b+c \text{ even, }a,b,c\geq1\} \\
    & \setminus \{XXX...X, \: YYY...Y, \: ZZZ...Z\} \Big )
\end{split}
\end{align}
%
All basis elements in this algebra commute with $XXX...X, \: YYY...Y,$ and therefore also commute with $ZZZ...Z$. Any other Pauli string apart from the ones in algebra does not commute either with $XXX...X$ or $YYY...Y$. 

In order to determine the dimension of $\galg(\text{Heisenberg})$, let us decompose $\mathfrak{su}(2^n) = \kalg \oplus \malg$ with $\theta(g) = XXX...X \: g \: XXX...X$. Then $\kalg$ is the subalgebra of $\mathfrak{su}(2^n)$ consisting of all the elements in $\mathfrak{su}(2^n)$ that commute with $XXX...X$. This decomposition of $\mathfrak{su}(2^n)$ is type A III, and resulting $\kalg$ is isomorphic to $\kalg \cong \mathfrak{su}(2^{n-1})\oplus \mathfrak{su}(2^{n-1}) \oplus \mathfrak{u}(1)$ \cite{dagli, drury2008quantum_shannon}. To have all the elements that commute both with $XXX...X$ and $YYY...Y$, let us further decompose $\kalg$ into $\kalg = \kalg' \oplus \malg'$ with the involution $\theta'(g) = YYY...Y \: g \: YYY...Y$. Therefore we have
\begin{align}
\begin{split}
    \kalg' = \mathrm{span}\{ \text{Pauli strings with } a \text{ many } X,\: b \text{ many } Y,\: c \text{ many } Z \big| a+b,a+c,b+c \text{ even, }a,b,c\geq1\}.
\end{split}
\end{align}
This decomposition does not affect $\mathfrak{u}(1)$ component of $\kalg$. It only affects $\mathfrak{su}(2^{n-1})$ pieces separately, and leads to $\kalg' \cong (\mathfrak{su}(2^{n-2})\oplus \mathfrak{su}(2^{n-2}) \oplus \mathfrak{u}(1)) \oplus (\mathfrak{su}(2^{n-2})\oplus \mathfrak{su}(2^{n-2}) \oplus \mathfrak{u}(1)) \oplus \mathfrak{u}(1)$. Therefore the dimension of $\kalg'$ is $|\kalg'|=4|\mathfrak{su}(2^{n-2})|+3=4^{n-1}-1$.

The difference between the basis of $\kalg'$ and $\galg(\text{Heisenberg})$ are the elements $XXX...X, \: YYY...Y,$ and $ZZZ...Z$. Therefore the dimension of the Heisenberg Hamiltonian algebra can be calculated as $|\galg(\text{Heisenberg})| = |\kalg'|-3 = 4^{n-1}-4$.

\section{Review of Involution}
\subsection{Involution and Cartan Decomposition}
\EK{
The Cartan decomposition is defined as the split given in Def. \EKK{1}, i.e. $\galg = \kalg \oplus \malg$. Finding such a split is difficult; assigning some basis elements into $\kalg$ and some into $\malg$ may lead to inconsistencies as we go through the basis elements of the full algebra $\galg$.} 

\EK{To get around this issue we may use
an involution $\theta$, i.e. a homomorphism on $\galg$ that preserves commutation relations and satisfies $\theta(\theta(g)) = g$ for all $g \in \galg$. This naturally
splits the algebra when one considers its $+1$ and $-1$ eigen-solutions (since the square of the involution is identity, $+1$ and $-1$ are the only possibilities for eigenvalues). Let us name $p_i$ as a $+1$ eigen-solutions, and $n_i$ as a $-1$ eigen-solutions where $i=1,2,3,...$ such that $\theta(p_i)=p_i$ and $\theta(n_i)=-n_i$. Then we can see that $[p_i,p_j]$ is a $+1$ eigen-solution:
\begin{align}
    \theta([p_i,p_j]) = [\theta(p_i),\theta(p_j)] = [p_i,p_j],
\end{align}
$[p_i,n_j]$ is a $-1$ eigen-solution:
\begin{align}
    \theta([p_i,n_j]) = [\theta(p_i),\theta(n_j)] = [p_i,-n_j] =  -[p_i,n_j],
\end{align}
and $[n_i,n_j]$ is a $+1$ eigen-solution:
\begin{align}
    \theta([n_i,n_j]) = [\theta(n_i),\theta(n_j)] = [-n_i,-n_j] =  [n_i,n_j].
\end{align}
Therefore if one defines $\kalg$ as the positive eigen-solution space for the involution $\theta$ and $\malg$ as the negative eigen-solution space, then the commutation relations given in Def. \EKK{1} are automatically satisfied.
}

\subsection{Involution Types for $\mathfrak{su}(N)$}
\EK{
The unitary operators involved in quantum
computing
fall into the Lie group $SU(N)$ which is generated by the Lie algebra $\mathfrak{su}(N)$, where $N=2^n$ and $n$ is the number of qubits. This algebra has infinitely many Cartan decompositions, many of them are equivalent to each other after a similarity transformation. However, there are 3 different classes of Cartan decompositions of $\mathfrak{su}(N)$ that cannot be transformed to each other which are named as AI, AII and AIII \cite{d2007introduction,quantum_symmetries}. Involutions that correspond to these (up to a similarity transformation) are as follows.
\begin{itemize}
    \item For AI, $\theta(g) = -g^T$ (In \cite{d2007introduction} this involution is given as complex conjugation rather than transpose, which is equivalent due to the fact that all the matrices we work are Hermitian matrices in this paper.)
    \item For AII (only when $N$ is even), $\theta(g) = -Mg^T M$, where
\begin{align}
    M = \begin{pmatrix}
        0 & -I_{N/2} \\
        I_{N/2} & 0
    \end{pmatrix}
\end{align}
and $I_{N/2}$ is the $N/2 \times N/2$ identity matrix
\item For AIII, $\theta(g) = U g U$ where $U$ is a diagonal matrix that has each diagonal element to be either $+1$ or $-1$ (note that $U^T=U$).
\end{itemize}
}

\subsection{Involution for Certain Models}\label{ycount_appendix}

Pauli matrices satisfy $X^T = X$, $Y^T = -Y$, $Z^T = Z$. Using $(A\otimes B)^T=A^T\otimes B^T$, one finds that Pauli strings with an even number of $Y$ matrices satisfy $\sigma^T = \sigma$, while the ones with an odd number of $Y$ matrices satisfy $\sigma^T = -\sigma$. Therefore, for any Lie subalgera $\galg \subseteq \mathfrak{su}(2^n)$ using the involution $\theta(g)=-g^T$
leads to a Cartan decomposition via $\theta(\kalg)=\kalg$, $\theta(\malg)=-\malg$ as
\begin{align}
\begin{split}
    \kalg &= \mathrm{span}\{\text{Pauli strings $\in \galg$ with odd $Y$ matrices}\}, \\
    \malg &= \mathrm{span}\{\text{Pauli strings $\in \galg$ with even $Y$ matrices}\}. 
\end{split}
\end{align}
The XY, transverse field XY, transverse field Ising and Heisenberg models have Hamiltonians consisting only of Pauli strings that have either $0$ or $2$ $Y$ matrices, therefore satisfy $\theta(\ham) = -\ham^T = -\ham$, which makes $\theta(g)=-g^T$ a suitable involution to apply Theorem \EKK{1} to these models.

\EK{
After the Jordan Wigner transformation, all these models fall into a fermion model that has time reversal symmetry. The reason that the involution $\theta(g) = -g^T$ work for these models can be explained by the relation between Cartan decompositions and discrete quantum symmetries as discussed in Ref.~\cite{quantum_symmetries}. There, it is explained that if the symmetry is due to a unitary transformation, then corresponding involution puts anything symmetric into $\kalg$. Parity in space can be considered as one of those. However, if the symmetry is due to an anti-unitary transformation such as time reversal, then the corresponding involution puts the symmetric element into $\malg$. Therefore it is guaranteed that there is an involution that will put $\ham$ into $\malg$ if the system has time reversal symmetry.     
}
\subsection{Involution Search}
\EK{
In our algorithm, we are limiting ourselves to single Pauli strings due to circuit composition considerations. This limits our involution pool for all AI, AII and AIII types, and leads to the following list:
\begin{enumerate}
    \item $\theta(g) = -B g^T B$ as AI, where $B$ is a Pauli string containing even number of $Y$ matrices,
    \item $\theta(g) = -B g^T B$ as AII, where $B$ is a Pauli string containing odd number of $Y$ matrices,
    \item $\theta(g) = B g B$ as AIII, where $B$ is a Pauli string.
\end{enumerate}
One can search for $B$ for any given Hamiltonian $\ham$ to find a suitable involution satisfying $\theta(\ham) = -\ham$.
}

\section{Method to find a Cartan Subalgebra}
\RTwo{
In this work, we specifically deal with basis elements that are single Pauli strings, i.e. tensor product of $n$ Pauli matrices where $n$ is the number of spins/qubits determined by the Hamiltonian. We exploit this fact while searching for a Cartan subalgebra. Suppose $\malg$ has the following basis
\begin{align}
    \malg = \mathrm{span} \{ \sigma_1, \sigma_2, \dots, \sigma_{|\malg|}\},
\end{align}
where each $\sigma_i$ is a Pauli string. We construct a list of basis elements for $\halg$ in the following way: we initialize a list of basis elements for $\halg$ via picking a random basis elements from $\malg$, say $\sigma_1$. Then we iterate through the basis elements of $\malg$ and append them if they commute with all the Pauli strings we have appended into the list. After a reordering the indices, we obtain the following $\halg$ without loss of generality
\begin{align}\label{aeq:halg}
    \halg = \mathrm{span} \{ \sigma_1, \sigma_2, \dots, \sigma_{|\halg|}\}.
\end{align}
With this notation, $\sigma_i \in \halg$ if $i \leq |\halg|$ and $\sigma_i \not\in \halg$ otherwise. 
\begin{theorem}\label{thm:maximal}
    The set given in Eq. \eqref{aeq:halg} is a maximal Abelian subalgebra of $\malg$.
\end{theorem}
}
\RTwo{
To prove this, we will prove the following lemma:
\begin{lemma}\label{lem:union}
    For $j = 1,2,\dots,|\halg|$, define the set of non-commuting indices as
    \begin{align}
        s(j) = \{  i \, | \: [\sigma_j, \sigma_i] \neq 0 \}
    \end{align}
    Then
    \begin{align}
        \bigcup_{j=1}^{|\halg|} s(j) = \{i\big|\:|\halg|<i\leq|\malg|\} = \{|\halg|+1,\dots,|\malg|\}.
    \end{align}
\end{lemma}
}
\RTwo{
\noindent \textbf{Proof:} For any $i,j\leq|\halg|$, we know that $[\sigma_j, \sigma_i] = 0$ because the condition to add the element in $\halg$ is that it commutes with the existing list of basis elements. This implies $i \not\in s(j)$ for any $i,j\leq|\halg|$, therefore the union of $s(j)$ sets cannot include any integer smaller than $|\halg|+1$. 
}
\RTwo{
For any $i>|\halg|$, we know that we could not add $\sigma_i$ into $\halg$ during the construction of $\halg$. This means that for any $|\halg|<i\leq|\malg|$, there exists a $j \leq |\halg|$ such that $[\sigma_j, \sigma_i] \neq 0$ i.e. $i \in s(j)$. Therefore union of all $s(1),s(2),\dots,s(|\halg|)$ must include all positive integers from $|\halg|+1$ to $|\malg|$. $\square$
}
\RTwo{
\begin{lemma}\label{lem:orthogonal}
    For $j = 1,2,\dots,|\halg|$ and $i,k = |\halg|+1,\dots,|\malg|$, if $[\sigma_j,\sigma_i]\neq0$ and $[\sigma_j,\sigma_k]\neq0$, then
    \begin{align}
        \tr\big([\sigma_j,\sigma_i][\sigma_j,\sigma_k]\big) = - 2^{n+2} \delta_{ik} 
    \end{align}
    where $\delta_{ik}$ is the Kronecker delta and $\sigma_i$ are $2^n \times 2^n$ matrices.
\end{lemma}
}
\RTwo{
\noindent \textbf{Proof:} Pauli matrices either commute or anti-commute. This extends to Pauli strings as well, because they are tensor products of Pauli matrices. Therefore if $[\sigma_j,\sigma_i]\neq0$ then $[\sigma_j,\sigma_i] = 2\sigma_j \sigma_i = - 2\sigma_i \sigma_j $. We further observe that
\begin{align}
    \tr\big([\sigma_j,\sigma_i][\sigma_j,\sigma_k]\big) &= 4 \tr\big(\sigma_j \sigma_i \sigma_j \sigma_k \big) \\
    &= - 4 \tr\big(\sigma_j \sigma_j \sigma_i \sigma_k \big) \\
    &= - 4 \tr\big(\sigma_i \sigma_k \big)
\end{align}
Now, if $i \neq k$ i.e. $\sigma_i \neq \sigma_k$, then $\sigma_i \sigma_k$ is not going to be identity: therefore it will be a tensor product of identity and at least one Pauli matrix-matrices. Since $\tr(A \otimes B) = \tr(A)\tr(B)$, we then have $\tr(\sigma_i \sigma_k) = 0$ if $i \neq k$. 
}
\RTwo{
In the $i = k$ case, $\tr(\sigma_i \sigma_k) = \tr(1) = 2^n$. Therefore
\begin{align}
    \tr\big([\sigma_j,\sigma_i][\sigma_j,\sigma_k]\big) = -4 \times 2^n \delta_{ik} = -2^{n+2} \delta_{ik},
\end{align}
which proves the lemma. $\square$.}
\\\\
\RTwo{
\noindent\textbf{Proof of Theorem \ref{thm:maximal}:} $\halg$ is an Abelian subalgebra of $\malg$ by construction. To prove maximality, we show that there is no other element in $\malg$, including linear combinations of Pauli strings, that commutes with all $\sigma_i$ for $i \leq |\halg|$. Consider a generic element $m = \sum_{i = 1}^{|\malg|} a_i \sigma_i \in \malg$. Then if
\begin{align}
    [\sigma_j,m] = \sum_{i = 1}^{|\malg|} a_i [\sigma_j,\sigma_i] = 0
\end{align}
for all $j \leq |\halg|$, then we get
\begin{align}
    \sum_{i \in s(j)} a_i [\sigma_j,\sigma_i] = 0 \:\:\:\:\text{ for }j=1,2,\dots,|\halg|.
\end{align}
Due to {orthogonality relations from} Lemma \ref{lem:orthogonal}, we then obtain $a_i = 0$ for all $i \in \cup_{j \leq |\halg|}s(j)$. Due to Lemma \ref{lem:union} this implies $a_i = 0$ for all $|\halg| < i \leq |\malg|$, which in turn results in
\begin{align}
    m = \sum_{i = 1}^{|\malg|} a_i \sigma_i = \sum_{i = 1}^{|\halg|} a_i \sigma_i \in \halg.  
\end{align}
Therefore we have proved that for any $m \in \malg$, if $[\halg,m] = 0$, then $m \in \halg$ which proves maximality. $\square$.
}

\section{Proof of the Theorem for Parameter Optimization}
\RTwo{
In this section, we will point out the differences between Theorem \redtext{2} and the methods given in \cite{helgason2001differential,earp2005constructive}. For this, let us start with stating the original KHK decomposition method as a theorem:
\begin{theorem}\label{thmA:earppachos}
\cite{helgason2001differential,earp2005constructive} For $\ham \in \malg$, define the function $f$
\begin{align}\label{eq:Earp_function_old}
    f(K) = \langle v,K^\dagger  \ham K \rangle =\langle K v K^\dagger, \ham \rangle,
\end{align}
where $\langle .,. \rangle$ denotes the Killing form, and $v \in \halg$ is an element whose exponential map $e^{itv}$ is dense in $e^{i \halg}$. Then for any {\em global minimum} of $f(K)$ denoted by $K_c$, 
\begin{align}
    K_c^\dagger \ham K_c \in \halg.
\end{align}
\end{theorem}
}

\RTwo{
Although it is stated that a global minimum is needed, both proofs given in \cite{helgason2001differential,earp2005constructive} only use the fact that the function $f$ has zero gradient at the global minimum, which makes the theorem work for any local extremum as well. By using this fact, we improve the theorem by showing that the element $K$ in the function $f$ does not need to be parameterized via exponential map as in \cite{khaneja2005constructive, earp2005constructive, drury2008quantum_shannon} ---  the parametrization/coordinate system $K = \exp(\sum_i \alpha_i k_i)$ is used to cover the entire Lie group $e^{i\kalg}$ which is not generally possible. Because we only need a local extremum, we can re-state the theorem for a generic parameterization system as the following:
}

\RTwo{
\begin{theorem}\label{thmA:earppachos_v2}
\textbf{(Improved KHK Decomposition)} Assume a set of coordinates $\Vec{\theta}$ in a chart of the Lie group $\exp(i \kalg)$.
For $\ham \in \malg$, define the function $f$
\begin{align}\label{eqA:Earp_function}
    f(\Vec{\theta}) = \langle K(\Vec{\theta}) v K(\Vec{\theta})^\dagger, \ham \rangle,
\end{align}
where $\langle .,. \rangle$ denotes \RThree{an invariant non-degenerate bilinear form on $\galg$}, and $v \in \halg$ is an element whose exponential map $e^{itv}$ is dense in $e^{i \halg}$. Then for any {\em local extrema} of $f(\Vec{\theta})$ denoted by $\Vec{\theta}_c$, and defining the critical group element $K_c = K(\Vec{\theta}_c)$, we have
\begin{align}
    K(\Vec{\theta}_c)^\dagger \ham K(\Vec{\theta}_c)= K_c^\dagger \ham K_c  \in \halg.
\end{align}
\end{theorem}
}

\RTwo{
The motivation for this extension is to use the decoupled product form (Eq. (\redtext{8}) in the main text) rather than the exponential map itself. To prove the theorem, we first provide the following definition.
}

\begin{definition}
    Given a compact Lie algebra $\kalg$ and Lie group $\kgroup = e^{i \kalg}$ generated via the exponential map. Let $f: \kgroup \to \mathbb{R}$ be a smooth function. Then, if for a coordinate system $\Vec{\theta}$ in $\kgroup$, partial derivatives of $f$ with respect to the coordinates vanish at $K_c \in \kgroup$ and the basis vectors at $K_c$ covers $\kalg$ i.e.
    \begin{align}
        &\frac{\partial f}{\partial \theta_i} \Big |_{K_c} = 0 \text{ for } i = 1,2,\dots, |\kalg|,  &\mathrm{span}\Big\{ K^\dagger \frac{\partial K}{\partial \theta_i} \Big |_{K_c} \Big\} = \kalg, 
    \end{align}
    then we will denote this critical point $K_c$ as a \textbf{non-singular} critical point in the coordinate system $\Vec{\theta}$. In the case the basis vectors not covering the $\kalg$, we will call it a \textbf{singular} critical point in the coordinate system $\Vec{\theta}$.
\end{definition}

\RTwo{
The product form (Eq. (\redtext{8})) might lead to coordinate singularities due to the fact that different set of coordinates might represent the same Lie group element. Therefore, if the product form is used in Theorem \ref{thmA:earppachos_v2}, local extremum might be a singular critical point. We provide the following lemma to cover this case as well. 
}

\RTwo{
\begin{lemma}\label{lem:criticality}
    If $K_c \in \kgroup$ is a singular critical point of the function $f:\kgroup \to \mathbb{R}$ in the coordinate system $\Vec{\theta}$, then there exists a coordinate system $\Vec{\alpha}$ such that $K_c$ is a non-singular critical point of $f$ in the coordinates $\Vec{\alpha}$.
\end{lemma}
}

\RTwo{
\noindent \textbf{Proof:}
We have 
\begin{align}
    &\frac{\partial f}{\partial \theta_i} \Big |_{K_c} = 0 \text{ for } i = 1,2,\dots, |\kalg|  &\text{span}\Big\{ K^\dagger \frac{\partial K}{\partial \theta_i} \Big |_{K_c} \Big\} \neq \kalg.
\end{align}
Since there are $|\kalg|$ many coordinates, some of the basis vectors should be linearly dependent because they fail to span $\kalg$. Let us say $r$ of them are linearly independent, where $r < |\kalg|$. Without loss of generality, we can choose them to be $\theta_1, \theta_2, ..., \theta_r$. 
We then conclude that there
are $r$ relevant coordinates that we denote as $\Vec{\eta}$,
and $|\kalg|-r$ irrelevant coordinates that we denote as $\Vec{\phi}$.
That is, \RTwo{denoting the coordinate mapping as $K(\eta,\phi) \in \kgroup$}, the Lie group element at the critical point $K_c \in e^{i\kalg}$ is determined by the relevant coordinates, and the irrelevant ones do not change the group element: $K_c = K(\Vec{\eta}_c,\Vec{\phi})$ for any choice of $\Vec{\phi}$. In other words, the map from this coordinate system to the Lie group manifold is many-to-one at the critical point $K_c$.
}
\RTwo{
\RTwo{We will exploit the fact that the function $f$ is}
a function of the group element $K$. Let us calculate the gradient of $f$ at $K_c$. \RTwo{Define a smooth curve $C(\lambda)$ on the manifold $C:\mathbb{R}^+ \to \kgroup$ such that the curve passed through at the critical point $C(\lambda = 0) = K_c$. Then gradient of $f$ on the direction of the curve $C$ at $\lambda = 0$ is given by 
\begin{align}\label{aeq:invariant}
    \frac{d}{d\lambda}f(C(\lambda)) \Big|_{\lambda = 0^+} = \lim_{\lambda \to 0^+} \frac{f(C(\lambda))-f(K_c)}{\lambda},
\end{align}
%
which is independent of coordinate choice. To write \eqref{aeq:invariant} in $\vec{\theta}$ coordinates, write $C(\lambda) = K(\vec{\eta}(\lambda),\vec{\phi}(\lambda))$. Then $C(0) = K_c$ implies that $\vec{\eta}(0) = \vec{\eta}_c$ while $\vec{\phi}(0)$ remains free since $K_c$ is independent of $\vec{\phi}$. This leads to
\begin{align}
    \frac{d}{d\lambda}f(C(\lambda)) \Big|_{\lambda = 0^+} = \lim_{\lambda \to 0^+} \frac{f(K(\vec{\eta}(\lambda),\vec{\phi}(\lambda)))-f(K(\vec{\eta}_c,\vec{\phi}))}{\lambda}.
\end{align}
Since $K_c$ does not depend on $\Vec{\phi}$, we can then write
\begin{align}
    \frac{d}{d\lambda}f(C(\lambda)) \Big|_{\lambda = 0^+} &= \lim_{\lambda \to 0^+} \frac{f(K(\vec{\eta}(\lambda),\vec{\phi}(\lambda)))-f(K(\vec{\eta}_c,\vec{\phi}(\lambda)))}{\lambda} \\
    &= \sum_{i=1}^r \frac{\partial f}{\partial \eta_i} \frac{d \eta_i}{d
    \lambda} \Big|_{\lambda = 0^+} = \sum_{i=1}^r \frac{\partial f}{\partial \theta_i} \frac{d \theta_i}{d
    \lambda} \Big|_{\lambda = 0^+} = 0
\end{align}
}
\RTwo{for any curve $C$. Let us choose a new set of coordinates}
\begin{align}
    \tilde{K}(\Vec{\alpha}) = K_c e^{i \sum_i \alpha_i k_i},
\end{align}
and rewrite \eqref{aeq:invariant} with $\Vec{\alpha}$ coordinates with $K_c = \tilde{K}(\vec{\alpha} = \vec{0})$ and $C(\lambda) = \tilde{K}(\vec{\alpha}(\lambda))$:
\RTwo{
\begin{align}
    0 = \frac{d}{d\lambda}f(C(\lambda)) \Big|_{\lambda = 0^+} &= \lim_{\lambda \to 0^+} \frac{f(\tilde{K}(\vec{\alpha}(\lambda)))-f(\tilde{K}(\vec{0}))}{\lambda} \\
    &= \sum_{i=1}^{|\kalg|} \frac{\partial f}{\partial \alpha_i} \frac{d \alpha_i}{d
    \lambda} \Big|_{\lambda = 0^+}.
\end{align}
Since the curve $C(\lambda)$ is arbitrary, $\frac{d \alpha_i}{d\lambda} \big|_{0^+}$ are arbitrary values, which yields}
\begin{align}
    \frac{\partial f}{\partial \alpha_i} \Big |_{K_c} = 0,
    \text{ for } i = 1,2,\dots, |\kalg|.
\end{align}
\RTwo{Therefore, $K_c$ is a critical point in the $\Vec{\alpha}$ coordinates as well.} 
Now let us show that the basis vectors generated by $\Vec{\alpha}$ coordinates will span $\kalg$. At $K = K_c$, one can easily see that 
\begin{align}
    K^\dagger \frac{\partial K}{\partial \alpha_i} \Big|_{K_c} = K_c^\dagger \lim_{\lambda \to 0} \frac{K_c e^{i \lambda k_i} - K_c}{\lambda} = k_i,
\end{align}
which yields
\begin{align}
    \text{span} \Big\{ K^\dagger \frac{\partial K}{\partial \alpha_i} \Big|_{K_c} \Big\} = \text{span} \Big\{ k_i \Big\} = \kalg
\end{align}
\RTwo{Therefore, $K_c$ is a non-singular critical point in the $\Vec{\alpha}$ coordinates, which proves the lemma.} $\square$
\\\\\\
\noindent \textbf{Proof of Theorem \ref{thmA:earppachos_v2}:} Let us calculate the partial derivatives of $f$ with respect to $\theta_i$: 
\begin{align}
\begin{split}
    \frac{\partial f(\Vec{\theta})}{\partial \theta_i} =& \left\langle \frac{\partial K}{\partial \theta_i} v K^\dagger, \ham \right\rangle + \left\langle K v \frac{\partial K^\dagger}{\partial \theta_i}, \ham \right\rangle \\
    =& \left\langle \frac{\partial K}{\partial \theta_i} v K^\dagger, \ham \right\rangle - \left\langle K v K^\dagger \frac{\partial K}{\partial \theta_i} K^\dagger , \ham \right\rangle \\
    =& \left\langle K K^\dagger \frac{\partial K}{\partial \theta_i} v K^\dagger, \ham \right\rangle - \left\langle K v K^\dagger \frac{\partial K}{\partial \theta_i} K^\dagger , \ham \right\rangle \\
    =& \left\langle K \Big[ K^\dagger \frac{\partial K}{\partial \theta_i}, v \Big] K^\dagger, \ham \right\rangle 
\end{split}
\end{align}
Using similarity transformation invariance of the \RThree{bilinear form}, we can shift $K (...) K^\dagger$ to the right
\begin{align}
\begin{split}
    \frac{\partial f(K)}{\partial \theta_i} =& \left\langle \Big[ K^\dagger \frac{\partial K}{\partial \theta_i}, v \Big], K^\dagger \ham K \right\rangle 
\end{split}
\end{align}
Now, we can rewrite this as 
\begin{align}\label{eq:extremum}
\begin{split}
    \frac{\partial f(K)}{\partial \theta_i} =& -i\frac{\partial}{\partial t}\left\langle e^{-itv}K^\dagger \frac{\partial K}{\partial \theta_i}e^{itv}, K^\dagger \ham K \right\rangle \Big|_{t=0} \\
    =& -i\frac{\partial}{\partial t}\left\langle K^\dagger \frac{\partial K}{\partial \theta_i},e^{itv} K^\dagger \ham K e^{-itv} \right\rangle  \Big|_{t=0} \\
    =& \left\langle K^\dagger \frac{\partial K}{\partial \theta_i},[v, K^\dagger \ham K] \right\rangle
\end{split}
\end{align}
Therefore, at the critical point $K = K_c$
\begin{align}
    \left\langle K^\dagger \frac{\partial K}{\partial \theta_i} \Big|_{K_c},[v, K_c^\dagger \ham K_c] \right\rangle = 0.
\end{align}
We know that $\ham, v \in \malg$ and $K_c \in e^{i\kalg}$. Therefore by the definition of Cartan decomposition (Def. \EKK{1} in the main text), $K_c \ham K_c^\dagger \in \malg$ and $[v, K_c^\dagger \ham K_c] \in \kalg$. \eqref{eq:extremum} is satisfied for all $i=1,2,\dots,|\kalg|$. {By the Lemma \ref{lem:criticality}, without loss of generality, we can simply assume that $K^\dagger \frac{\partial K}{\partial \theta_i} \Big|_{K_c}$ span $\kalg$}. \RThree{Then due to non-degeneracy of the bilinear form,} \eqref{eq:extremum} yields  
\begin{align}
 [v, K_c^\dagger \ham K_c] = 0
\end{align}
which also means
\begin{align}\label{eq:exponentialdense}
 [e^{itv}, K_c^\dagger \ham K_c] = 0.
\end{align}
The exponential map of $v$ is dense in $e^{i\halg}$, i.e.
for any element $e^{ih}$ chosen in the group $e^{i\halg}$, the line $e^{itv}$ passes through a point that is arbitrarily close to the element $e^{ih}$. This with \eqref{eq:exponentialdense} means that $K^\dagger \ham K$ commutes with any element in $\halg$. Since $\halg$ is the maximal Abelian Lie algebra in $\malg$, this implies that $K^\dagger \ham K \in \halg$. $\square$.
}

\RTwo{
We then conclude that the method given in Refs. \cite{helgason2001differential,earp2005constructive} \textbf{does not} require us to minimize the function $f$. It requires us to find a \textbf{local} extremum. In addition, this gives us flexibility in how to represent the element $K$ in Theorem \ref{thmA:earppachos_v2}. This additional property will be exploited in the next section.
}

\section{Product Anzats for $K$}
\RTwo{
In contrast to other works using Cartan decomposition \cite{khaneja2005constructive, earp2005constructive, drury2008quantum_shannon} that represent an element $K \in e^{i\kalg}$ as $K = \exp(\sum_i \alpha_i k_i)$, where $k_i$ is the $i$th basis element of $\kalg$, we use the following product representation:
\begin{align}
    K(\Vec{\theta}) = \prod_{i} e^{i \theta_i k_i},
    \label{aeq:kform}
\end{align}
For example, for $\kalg = \mathfrak{su}(2)$, this product would be
\begin{align}
    K(a,b,c) = e^{i a X}e^{i b Y}e^{i c Z}.
\end{align}
}

\RTwo{
The product expansion \eqref{aeq:kform} is extremely beneficial when the basis elements $k_i$ are Pauli strings, because it can then be directly implemented in a \RThree{quantum computer} and we only need to find the parameters $\theta_i$ to generate the circuit. In \cite{khaneja2005constructive}, Cartan decomposition is applied recursively because the expression $K = \exp(\sum_i \alpha_i k_i)$ cannot be implemented on a quantum device, which is the unitary synthesis problem that we try to solve for Hamiltonian evolution in the first place. A second benefit of this expression comes from the optimization: as explained in the next section, \eqref{aeq:kform} allows us to calculate $f(K)$ and its gradient with a greater accuracy.
}

\RTwo{
{The product expansion \eqref{aeq:kform} does not cover the Lie group $e^{i\kalg}$ except special cases such as $\kalg$ being Abelian or solvable \cite{wei_norman}. However, it is a good parametrization that can be used in the theorem. To show that, let us first notice that the \eqref{aeq:kform} is differentiable.} Secondly, let us show that \eqref{aeq:kform} covers a $|\kalg|$ dimensional subspace of $e^{i\kalg}$. For this, observe that near identity $K(\Vec{\theta}=0)=I$ \eqref{aeq:kform} can be expanded as
\begin{align}
    K(\Vec{\theta}) = I + \sum_{i} \theta_i k_i + O(\theta^2),
    \label{aeq:kform_near_identity}
\end{align}
Since each $k_i$ are basis elements of $\kalg$, it is obvious that the equation above covers $|\kalg|$ dimensional neighborhood of the identity element. Considering the continuity and the differentiability of the parametrization \eqref{aeq:kform}, we can therefore easily conclude that it covers a $|\kalg|$ dimensional subspace of the Lie group $e^{i\kalg}$. Therefore this parametrization can be used in Theorem \ref{thmA:earppachos_v2}.
}
\RTwo{
Now let us show that the $f(\Vec{\theta}) = f(K(\Vec{\theta}))$ has a local extremum. For our specific case, we pick our basis elements from single Pauli strings. In this case
\begin{equation}
    e^{i \theta_i k_i} = \cos \theta_i I + i \sin \theta_i \: k_i,
\end{equation}
Therefore $f(\Vec{\theta})$ is a periodic function on all its variables. Since it is also differentiable, it should have a local extremum within a period. Therefore by using the product expansion, we are guaranteed to find a local extremum and therefore find a solution to our decomposition.
}

\section{Time Complexity for Parameter Optimization}\label{func_appendix}

\subsection{Cost function evaluation}

To perform the optimization, we need to calculate $f(K) = \big< K v K^\dagger, \ham \big>$ \RThree{given in Theorem \ref{thmA:earppachos_v2} (Theorem \redtext{2} in the main text)} where $\ham \in \malg$, $K\in e^{i\kalg}$ and $v$ is an element in $\halg$ whose exponential map $e^{itv}$ is dense in $e^{i\halg}$. \RTwo{ $\halg$ is an Abelian Lie algebra and in this work, basis elements of $\halg$ are single Pauli strings. Therefore the parameter space for the group $e^{i\halg}$ is $2\pi$ periodic on all parameters, meaning that it is a $|\halg|$ dimensional torus. If one chooses $v$ as $a_1 h_1 + a_2 h_2 + ...$ where $a_i$ are mutually irrational to each other, then the line $e^{itv}$ will be dense in $e^{i\halg}$. Therefore}
we use 
$v = \sum_i \gamma^i h_i$
%
where $\gamma$ is a transcendental number \EK{to ensure that any power $\gamma^i$ is irrational.}

We represent $K$ with the following product of exponentials
\begin{align}\label{eq:K_product_appendix}
    K = \prod_i e^{i \theta_i k_i},
\end{align}
where $k_i$ form a basis
for $\mathfrak{k}$. Using the fact that Killing form $\big<A,B\big>$ in $\mathfrak{su}(2^n)$ is proportional to $\tr(AB)$ where $\tr$ is the matrix trace, \RThree{and therefore it is a non-degenerate invariant bilinear form in $\galg(\ham) \subset \mathfrak{su}(2^n)$, we can replace $\langle A,B \rangle$ in the function with $\tr(AB)$. Then}
we find
\begin{align}\label{eq:f_appendix}
    f(K) = \tr \Big( \prod_{i \uparrow} e^{i \theta_i k_i} \: v \: \prod_{i \downarrow} e^{-i \theta_i k_i} \: \ham  \Big),
\end{align}
where $\uparrow$ ($\downarrow$) under the product means multiplication in an increasing (decreasing) order for $i$. Efficient calculation of this product is non-trivial because the products of these exponentials generally will not be in $\galg$, and thus may generate an arbitrary matrix in $\mathrm{GL}(2^n)$ and therefore require an exponential amount of calculation. One fact that can be used is that if $K \in e^{i\kalg}$ and $m \in \malg$, then we have $KmK^\dagger \in \malg$. Thus, if each exponential on both sides of $v$ in \eqref{eq:f_appendix} is applied on $v$, one from each side at the same time (a similarity transformation), the result will always be in $\malg$. To apply the exponentials of $k_i$, one can take advantage of the fact that $k_i$ are Pauli strings, therefore $k_i^2 = I$ and 
\begin{align}\label{eq:expPauli_appendix}
    e^{i \theta_i k_i} = \cos\theta_i \: I + i \sin\theta_i \: k_i. 
\end{align}
This allows us to apply one similarity transformation on one term in $v$ via constant amount of calculations only requires a constant amount of calculations.
After applying all exponentials to $v$, 
\begin{align}
    f(K) = \tr \Big( m_0 \ham  \Big)
\end{align}
is obtained, where
\begin{align}
    m_0 = \prod_{i \uparrow} e^{i \theta_i k_i} \: v \: \prod_{i \downarrow} e^{-i \theta_i k_i} \in \malg.
\end{align}
Up to this point, in the worst case only $\bigO(|\kalg||\malg|)$ many operations are performed.
$|\kalg|$-many exponentials are applied to an element of $\malg$, which has at most $|\malg|$ many Pauli terms. Multiplying $m_0$ and $\ham$ requires $\bigO(|\malg|)$ many calculations and can be neglected in the $|\kalg|\gg 1$ limit corresponding to large system size limit, which leads to $\bigO(|\kalg||\malg|)$ time complexity to calculate $f(K)$.


\subsection{Gradient evaluation}\label{grad_appendix}

The gradient of $K$ given in \eqref{eq:K_product_appendix} can be expressed as
\begin{align}\label{eq:K_derivative}
   \frac{\partial K}{\partial \theta_j}  = \prod_{i<j, \uparrow} e^{i \theta_{i} k_i} i k_j \prod_{i\geq j, \uparrow} e^{i \theta_{i} k_i}, 
\end{align}
leading to the analytical expression for the gradient of the function \eqref{eq:f_appendix}:

\begin{align}\label{eq:gradf_appendix}
\begin{split}
    \frac{\partial f(K)}{\partial \theta_j} =& \tr \Big( \prod_{i<j, \uparrow} e^{i \theta_{i} k_i} i k_j \prod_{i\geq j, \uparrow} e^{i \theta_{i} k_i} \: v \: \prod_{i \downarrow} e^{-i \theta_i k_i} \: \ham  \Big)\\
    + &\tr \Big( \prod_{i \uparrow} e^{i \theta_{i} k_i} \: v \: \prod_{i\geq j, \downarrow} e^{i \theta_{i} k_i} (-i)k_j \prod_{i<j, \downarrow} e^{-i \theta_{i} k_i} \: \ham  \Big).
\end{split}
\end{align}
Using the cyclic property of trace $\tr(AB)=\tr(BA)$ leads to  
\begin{align}
\begin{split}
    \frac{\partial f(K)}{\partial \theta_j} =& i \tr \Big( k_j \prod_{i\geq j, \uparrow} e^{i \theta_{i} k_i} \: v \: \prod_{i \downarrow} e^{-i \theta_i k_i} \: \ham \prod_{i<j, \uparrow} e^{i \theta_{i} k_i}  \Big)\\
    -i &\tr \Big( \prod_{i<j, \downarrow} e^{-i \theta_{i} k_i} \: \ham \: \prod_{i \uparrow} e^{i \theta_{i} k_i} \: v \: \prod_{i\geq j, \downarrow} e^{i \theta_{i} k_i} \: k_j    \Big)
\end{split}
\end{align}
Applying the exponentials one by one from both sides of $v$ and $\ham$, we obtain 
\begin{align}
    \frac{\partial f(K)}{\partial \theta_j} = i \tr \Big( k_j \: m_1 \: e^{-i \theta_j k_j} \: m_2  \Big)-i\tr \Big( m_2 \: e^{i \theta_j k_j} \: m_1 \: k_j    \Big).
\end{align}
where
\begin{align}
\begin{split}
    m_1 &= \prod_{i\geq j, \uparrow} e^{i \theta_{i} k_i} \: v \: \prod_{i \geq j \downarrow} e^{-i \theta_i k_i} \in \malg \\
    m_2 &= \prod_{i<j \downarrow} e^{-i \theta_i k_i} \: \ham \prod_{i<j, \uparrow} e^{i \theta_{i} k_i} \in \malg.
\end{split}
\end{align}
As in the calculation of $f(K)$, reaching that point costs $\bigO(|\kalg||\malg|)$ amount of time and in the $|\kalg|\gg 1$ limit corresponding to large system size limit, it is the most time consuming part compared to the last calculation of trace which takes $\bigO(|\malg|)$ time as above. However, this complexity is to obtain just one derivative. To calculate the full gradient, one has to perform this for all $\theta_j$, and therefore the complexity of calculating the entire gradient is $\bigO(|\kalg|^2|\malg|)$. 

\subsection{Obtaining $h \in \halg$}
\RTwo{After finding $K_c\in e^{i\kalg}$ that locally extremizes $f(K)$, one can obtain $h\in\halg$ via the following (also Eq. (\redtext{7}) in the main text):
\begin{align}
    h = K_c^\dagger \ham K_c \in \halg.
\end{align}
Using the product form of $K_c$, one finds
\begin{align}
    h = \prod_{i \downarrow} e^{-i \theta_i k_i} \: \ham \: \prod_{i \uparrow} e^{i \theta_i k_i}.
\end{align}
As discussed in the previous sections, the similarity transformations from each single Pauli exponential $e^{i \theta_i k_i}$ can be applied one by one with analytical precision via \eqref{eq:expPauli_appendix}, and since $\ham \in \malg$, the terms generated after each similarity transformation are also in $\malg$. This leads to $\bigO(|\kalg||\malg|)$ complexity, just as for the calculation of $f(K)$.}

\section{2 site TFIM Parameter Fit}

As given in Fig. \EKK{2}, the 2 site transverse field Ising model, $\ham = ZZ + B_1 IX + B_2 XI$, has the following Hamiltonian algebra
\begin{align}
    \galg(\ham) = \mathrm{span}\{XI,IX,ZZ,YY,YZ,ZY \},
\end{align}
and the following Cartan decomposition and Cartan subalgebra are used
\begin{align}
\begin{split}
    \kalg &= \mathrm{span}\{YZ,ZY \}, \\
    \malg &= \mathrm{span}\{XI,IX,ZZ,YY \}, \\
    \halg &= \mathrm{span}\{XI,IX \}.
\end{split}
\end{align}
By defining $v = IX + \gamma XI$, with $\gamma$ an arbitrary
transcendental constant,
and $K = e^{iaYZ} e^{ibZY}$, the cost function (\EKK{7}) can be calculated as
\begin{align}
\begin{split}
    f(a,b) =& \tr\big(e^{iaYZ}e^{ibZY} (IX + \gamma XI) e^{-ibZY}e^{-iaYZ} \ham \big) \\
    =&(B_1 + \gamma B_2)\cos 2a \cos 2b - (B_2 + \gamma B_1) \sin 2a \sin 2b + \cos 2a \sin 2b + \gamma \sin 2a \cos 2b.
\end{split}
\end{align}
To find a local extremum, we set $\partial f/\partial a = \partial f/\partial b = 0$, which yields 
%
\begin{align}
\begin{split}
    \tan(2a+2b) &= \frac{1}{B_1+B_2}, \\
    \tan(2a-2b) &= \frac{1}{B_2-B_1},
\end{split}
\end{align}
and are solved by 
\begin{align}
\begin{split}
    a &= \frac{1}{4}\arctan \Big(\frac{1}{B_1+B_2}\Big)-\frac{1}{4}\arctan \Big(\frac{1}{B_1-B_2}\Big), \\
    b &= \frac{1}{4}\arctan \Big(\frac{1}{B_1+B_2}\Big)+\frac{1}{4}\arctan \Big(\frac{1}{B_1-B_2}\Big).
\end{split}
\end{align}
Plugging this in $K^\dagger \ham K$, one finds
\begin{align}
\begin{split}
    K^\dagger\ham K =& e^{-ibZY}e^{-iaYZ} \ham e^{iaYZ}e^{ibZY} \\
    =& IX \Big( \frac{(B_1+B_2)^2-1}{2 \sqrt{1+(B_1+B_2)^2}} - \frac{(B_1-B_2)^2-1}{2 \sqrt{1+(B_1-B_2)^2}}  \Big) \\
    +& XI \Big( \frac{(B_1+B_2)^2-1}{2 \sqrt{1+(B_1+B_2)^2}} + \frac{(B_1-B_2)^2-1}{2 \sqrt{1+(B_1-B_2)^2}}  \Big) \\
    =& c \: IX + d \: XI\in \halg.
\end{split}
\end{align}
With this, we have $\mathcal{H}=K(cIX+dXI)K^\dagger$, which is the desired relationship.

\section{Circuit Optimization for TFXY Model}

In this section, we outline several circuit optimizations
which we apply to the free-fermionizable model discussed in the
main text. 

Consider an $n$-qubit circuit. We establish the following Lemma:
\begin{lemma}\label{thm:rocket}
    For any $i,j = 1,2,...,n-1$, $i<j$ and any $\alpha,\beta\in \mathbb{R}$, there exist $a,b,c \in \mathbb{R}$ such that 
    \begin{align}\label{eq:Euler}
        e^{i\alpha\widehat{Y_i X}_j}e^{i\beta\widehat{Y_i X}_{j+1}} = e^{ia\widehat{Y_j X}_{j+1}}e^{ib\widehat{Y_i X}_j}e^{ic\widehat{Y_j X}_{j+1}} 
    \end{align}
    where the ``hat" notation is defined in Eq. \EKK{12}. The same is true for $X \leftrightarrow Y$.
\end{lemma}

To prove this, observe that the algebra generated by the exponents of the left hand side is a representation of $\mathfrak{su}(2)$:
\begin{align}
\begin{split}
    [\widehat{Y_i X}_{j+1},\widehat{Y_i X}_j] &= 2i \: \EKK{\widehat{Y_j X}_{j+1}} \\
    [\EKK{\widehat{Y_j X}_{j+1}},\widehat{Y_i X}_{j+1}] &= 2i \: \widehat{Y_i X}_j \\
    [\widehat{Y_i X}_j,\EKK{\widehat{Y_j X}_{j+1}}] &= 2i \: \widehat{Y_i X}_{j+1}
\end{split}
\end{align}
Thus, \eqref{eq:Euler} is an Euler decomposition of a $\mathfrak{su}(2)$ spanned by the Pauli strings. The version with  $X \leftrightarrow Y$ is also true for the same reason.

\begin{figure}[htpb]

    \includegraphics[width = 0.25\columnwidth]{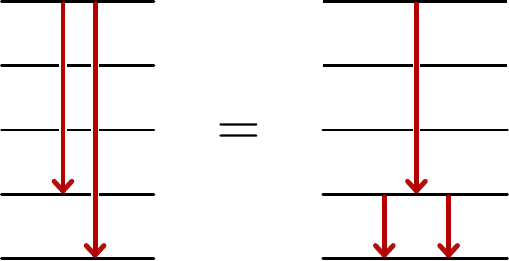}

    \caption{Graphical representation of Lemma \ref{thm:rocket} for $i=1,j=4$.}
    \label{fig:rocket}
\end{figure}

\begin{figure}[htpb]

    \includegraphics[width = 0.5\columnwidth]{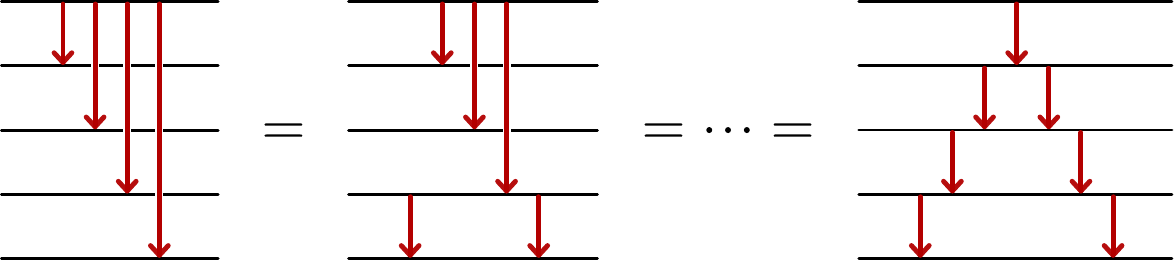}

    \caption{Graphical representation of Theorem \ref{thm:recursive} for $i=5$.}
    \label{fig:zigzag}
\end{figure}

Using the arrow notation introduced in Fig. \EKK{3}, a graphical representation of Lemma \ref{thm:rocket} can be given as Fig.~\ref{fig:rocket}. Next, we apply this Lemma to an ordered product of exponentials
as is used in Eqs.~\eqref{eq:f_appendix} and \eqref{eq:K_derivative}.
\begin{theorem}\label{thm:recursive}
    We define a ``triangle'' of size $i$ as 
    \begin{align}
        T_i(\vec \alpha) = \prod_{j=1}^i e^{i \alpha_j \widehat{Y_1 X}_j}
    \end{align}
    and
    \begin{align}
        Z_{p,q}(\vec \alpha) = \prod_{j=p}^q e^{i \alpha_j \widehat{Y_j X}_{j+1}}
    \end{align}
    which will be denoted as ``zig'' if $p>q$ and ``zag'' if $p<q$. Then for $i \geq 3$, there exists a new set of parameters $\vec a,b,c \in \mathbb{R}$ such that  
    \begin{align}\label{eq:recursion}
        T_i(\vec\alpha) = e^{ib \widehat{Y_{j-1} X}_j } T_{i-1}(\vec a)  e^{ic \widehat{Y_{j-1} X}_j }
    \end{align}
    and this implies for a new set of parameters $\vec\beta, \vec\theta \in \mathbb{R}$, the ``triangle'' $T_i(\vec \alpha)$ can be written as a ``zigzag"
    \begin{align}\label{eq:zigzag}
    \begin{split}
        T_i(\vec\alpha) &= \prod_{j=i-1, \downarrow}^1 e^{i \beta_j \widehat{Y_j X}_{j+1}}\prod_{j=1}^{i-1} e^{i \theta_j \widehat{Y_j X}_{j+1}} \\
        &= Z_{i-1,1}(\vec\beta)Z_{2,i-1}(\vec\theta).
    \end{split}
    \end{align}
\end{theorem}
To prove this, we first observe that
\begin{align}
\begin{split}
    T_i(\vec\alpha) =& \prod_{j=1}^i e^{i \alpha_j \widehat{Y_1 X}_j} \\
    =& \Bigg( \prod_{j=1}^{i-2} e^{i \alpha_j \widehat{Y_1 X}_j} \Bigg)e^{i \alpha_{i-1} \widehat{Y_1 X}_{i-1}}e^{i \alpha_i \widehat{Y_1 X}_i}.
\end{split}
\end{align}
Using Lemma \ref{thm:rocket} on the last two exponentials, and renaming new parameters:
\begin{align}
\begin{split}
    T_i(\vec \alpha) 
    &= \Bigg( \prod_{j=1}^{i-2} e^{i a_j \widehat{Y_1 X}_j} \Bigg)e^{i b \widehat{Y_{i-1} X}_{i}}e^{i a_{i-1} \widehat{Y_1 X}_{i-1}}e^{i c \widehat{Y_{i-1} X}_i} \\
    &= e^{i b \widehat{Y_{i-1} X}_{i}}\Bigg( \prod_{j=1}^{i-2} e^{i a_j \widehat{Y_1 X}_j} \Bigg)e^{i a_{i-1} \widehat{Y_1 X}_{i-1}}e^{i c \widehat{Y_{i-1} X}_i}\\
    &= e^{i b \widehat{Y_{i-1} X}_{i}}\Bigg( \prod_{j=1}^{i-1} e^{i a_j \widehat{Y_1 X}_j} \Bigg)e^{i c \widehat{Y_{i-1} X}_i}\\
    &= e^{i b \widehat{Y_{i-1} X}_{i}}T_{i-1}(\vec a)e^{i c \widehat{Y_{i-1} X}_i},
\end{split}
\end{align}
which is just equation \eqref{eq:recursion}. Recursively iterating this a total of $i-1$ times, results in Eq.~\eqref{eq:zigzag}. In the graphical representation, this recursion is easy to see as shown in Fig.~\ref{fig:zigzag}.

\begin{figure}[htpb]

    \includegraphics[width = 0.5\columnwidth]{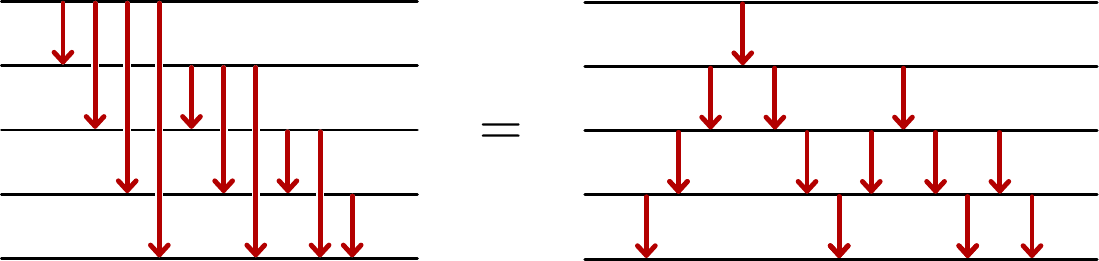}
    
    \caption{First simplification of the initial circuit for $K$.}
    \label{fig:first_simp}
\end{figure}

Considering the fact that the original circuit is a product of triangles, we now show that it can be written as a series of zigzags schematically depicted in Fig.~\ref{fig:first_simp}. This greatly reduces the complexity of the circuit because the initial circuit given on the left has $\bigO(n^3)$ CNOT gates, whereas the simplified zigzag circuit has only $\bigO(n^2)$ CNOT gates. However, this circuit can be simplified further.

\begin{lemma}\label{thm:flip}
    Any zigzag can be flipped into a ``zagzig", i.e. for any $i \geq 2$ and any set of parameters $\alpha_j, \beta_j \in \mathbb{R}$, there exists $\vec a,\vec b \in \mathbb{R}$ such that
    \begin{align}
        Z_{i,1}(\vec \alpha)Z_{2,i}(\vec \beta) = Z_{1,i}(\vec a)Z_{i-1,1}(\vec b)
    \end{align}
\end{lemma}

\begin{figure}[htpb]

    \includegraphics[width = 0.35\columnwidth]{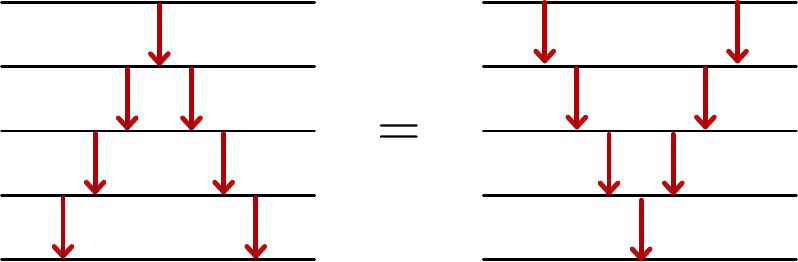}
    \caption{Graphical representation of Lemma \ref{thm:flip} for $i=5$.}
    \label{fig:flip}
\end{figure}

The proof is by induction. The base case is for zigzags with size $i=2$, because $\widehat{Y_2 X}_{3},\widehat{Y_1 X}_{2}, \widehat{Y_1 X}_{3}$ forms a representation of $\mathfrak{su}(2)$ and it is established by using the  Euler decomposition in the following two ways:
\begin{align}\label{eq:basecase}
    e^{i a \widehat{Y_1 X}_{2}}e^{i b \widehat{Y_2 X}_{3}} 
    e^{i c \widehat{Y_1 X}_{2}} = e^{i \alpha \widehat{Y_2 X}_{3}}e^{i \beta \widehat{Y_1 X}_{2}} 
    e^{i \theta \widehat{Y_2 X}_{3}}.
\end{align}
This is precisely the $i=2$ case of the lemma. Now we assume that it also holds for all zigzags up to size $N$. Then for $i=N+1$, we have that
\begin{align}
\begin{split}
    Z_{N+1,1}(\vec \alpha)Z_{2,N+1}(\vec \beta)
    =& Z_{N+1,3}(\vec \alpha) 
    \Big( e^{i \alpha_2 \widehat{Y_2 X}_{3}}e^{i \alpha_1 \widehat{Y_1 X}_{2}}e^{i \beta_2 \widehat{Y_2 X}_{3}} \Big)
    Z_{3,N+1}(\vec \beta)
\end{split}
\end{align}
The product in parentheses is our base case shown in Eq.~\eqref{eq:basecase}. Therefore, for some $a,b,c \in \mathbb{R}$, we have that
\begin{align}
\begin{split}
    Z_{N+1,1}(\vec \alpha)Z_{2,N+1}(\vec \beta)
    =& Z_{N+1,3}(\vec \alpha) 
    \Big( e^{i a \widehat{Y_1 X}_{2}}e^{i b \widehat{Y_2 X}_{3}} 
    e^{i c \widehat{Y_1 X}_{2}} \Big)
    Z_{3,N+1}(\vec \beta)\\
    =& e^{i a \widehat{Y_1 X}_{2}} \Big(Z_{N+1,3}(\vec \alpha) \:\:
    e^{i b \widehat{Y_2 X}_{3}}
    Z_{3,N+1}(\vec \beta) \Big) e^{i c \widehat{Y_1 X}_{2}}.
\end{split}
\end{align}
Note that the expression inside the parentheses on the last line is a zigzag with size $N$, that runs between sites $2$ and $N+1$. Therefore it can be flipped by the induction hypothesis. After renaming the parameters $a \to a_1$ and $c \to c_1$, we obtain
\begin{align}
\begin{split}
    Z_{N+1,1}(\vec \alpha)Z_{2,N+1}(\vec \beta)
    =& e^{i a_1 \widehat{Y_1 X}_{2}} \Big(Z_{2,N+1}(\vec a) Z_{N,2}(\vec b) \Big) e^{i c \widehat{Y_1 X}_{2}}\\
    =& Z_{1,N+1}(\vec a)Z_{N,1}(\vec b),
\end{split}
\end{align}
which proves the induction step. This completes the proof of Lemma \ref{thm:flip}.
A graphical representation is given in Fig.~\ref{fig:flip}.

\begin{theorem}\label{thm:pile}
    The zigzag circuit given for $K$ can be simplified into a more compact multiplications of ``zigs", i.e. for every $n \geq m+1$, $\vec \alpha_i, \vec \beta_i \in \mathbb{R}$ there exist a set of $\vec \theta_i \in \mathbb{R}$, such that
    \begin{align}
        \prod_{i=m}^{n-1}\big( Z_{n,i}(\vec \alpha_{i})Z_{i+1,n}(\vec \beta_{i})\big) = \prod_{i=m}^{n} Z_{n,i}(\vec \theta_{i}).
    \end{align}
\end{theorem}

For convenience, the parameters will not be shown explicitly for this proof i.e. $Z_{n,m}(\vec \alpha)$ will be written as $Z_{n,m}$, since the parameters are not determined explicitly in the argument. The proof is again by induction. The base case is for $n-m=1$ since both sides become $Z_{m+1,m} Z_{m,m}$. Now, assume that the induction step holds  for all $n-m$ up to $n-m=N \geq 1$. We will next establish that it holds for $n-m=N+1$. First, define $N' = N+m+1$ and then regroup the product to obtain
\begin{align}
\begin{split}
    \prod_{i=m}^{N'-1}\big( Z_{N',i} Z_{i+1,N'}\big) = Z_{N',m} \prod_{i=m+1}^{N'-1}\big( Z_{i,N'}Z_{N',i}\big)Z_{N',N'}.
\end{split}
\end{align}
Since $N'>i$, $\:Z_{i,N'}Z_{N',i}$, the product of terms in the parenthesis can be rewritten as $Z_{i,N'}Z_{N'-1,i}$, yielding
\begin{align}
\begin{split}
    = Z_{N',m} \prod_{i=m+1}^{N'-1}\big( Z_{i,N'}Z_{N'-1,i}\big)Z_{N',N'}.
\end{split}
\end{align}
Using Lemma \ref{thm:flip} for the expression in the product, we find that the product becomes
\begin{align}
\begin{split} 
    =& Z_{N',m} \prod_{i=m+1}^{N'-1}\big( Z_{N',i}Z_{i+1,N'}\big)Z_{N',N'}\\
    =& Z_{N',m} \prod_{i=m+1}^{N'-1}\big( Z_{N',i}Z_{i,N'}\big).
\end{split}
\end{align}
In the last step, we used the fact that the last term in the product in the middle is  $Z_{N',N'}$ and therefore can be absorbed into the term after the product by redefining its coefficient in the exponent. 

\begin{figure}[htpb]

    \includegraphics[width = 0.5\columnwidth]{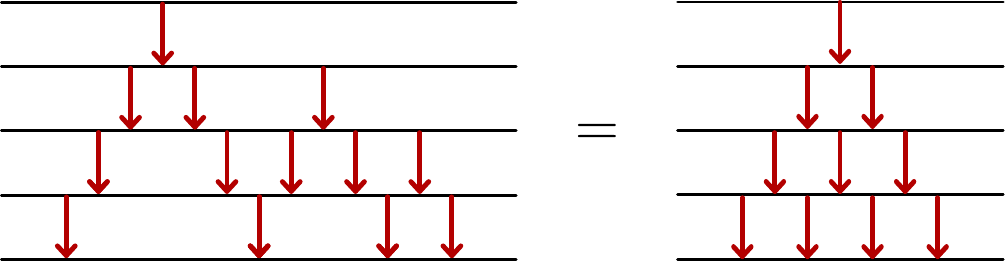}
    \caption{Graphical representation of theorem \ref{thm:pile} for $n=4$ and $m=1$.}
    \label{fig:pile}
\end{figure}

Note that the product term to the right is part of the induction hypothesis for $n-m=N$. Applying the induction hypothesis gives us
\begin{align}
\begin{split} 
    =& Z_{N',m} \prod_{i=m+1}^{N'}\big( Z_{N',i}\big) \\
    =& \prod_{i=m}^{N'}\big( Z_{N',i}\big),
\end{split}
\end{align}
which proves the induction step, and establishes the theorem. A graphical representation is given in Fig.~\ref{fig:pile}.

\begin{figure}[htpb]

    \includegraphics[width = 0.7\columnwidth]{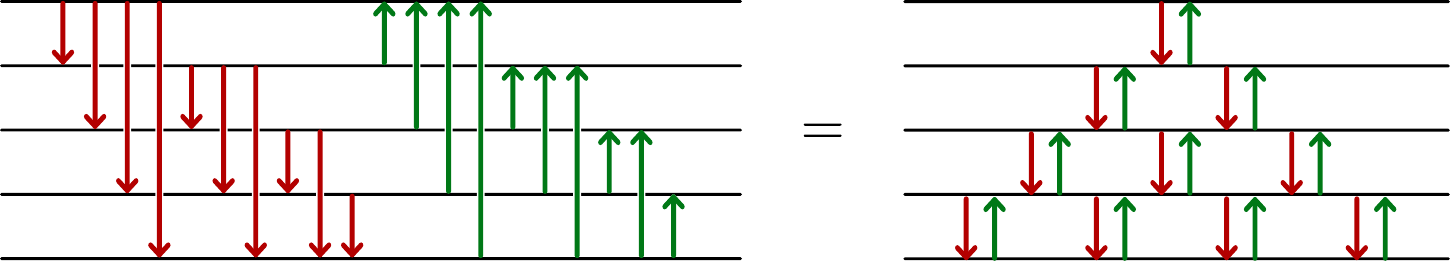}
    \caption{Simplification of K circuit for Transverse Field XY model for 5 spins.}
    \label{fig:simple_K}
\end{figure}

Using these results, we find that the red part of the original K circuit, given in the left side of Fig.~\ref{fig:first_simp}. can be rewritten as the circuit on right shown in Fig.~\ref{fig:flip}. Considering that all the down red arrows commute with all the up green arrows, we can move greens through the reds and arrive at the  simplification shown in Fig.~\ref{fig:simple_K}.

As shown in Fig. \EKK{3}(a), an arrow with length $a$, \textit{i.~e.}, the circuit for $\text{exp}(i\theta \widehat{X_i Y}_{i+a})$ (or the same circuit, but with $X \leftrightarrow Y$), has $2a$ CNOT gates in it. Therefore, the number of CNOT gates in the circuit on the left of Fig.~\ref{fig:simple_K} satisfies
\begin{align}
\begin{split}
    \text{\#CNOTs for raw K} = \text{(red part)} + \text{(green part)} = 2\sum_{p=1}^{n-1} \sum_{q=1}^{p} (2q) = 2\sum_{p=1}^{n-1} p(p+1) = \frac{2n(n^2-1)}{3}.
\end{split}    
\end{align}
On the other hand, the optimized circuit for $n$ spins consists pairs of length one red arrows followed by length one green arrows, that is $\text{exp}(i\theta \widehat{Y_i X}_{i+1})\:\text{exp}(i\phi \widehat{X_i Y}_{i+1})$. A circuit for this pair requires only 2 CNOTs \cite{vidal2004universal}. Therefore, the total CNOT count of the simplified circuit on the left of Fig.~\ref{fig:simple_K} is reduced to only the following:
\begin{align}
\begin{split}
    \text{\#CNOTs for simplified K} = \sum_{p=1}^{n-1} (2p) = n(n-1).
\end{split}    
\end{align}
The full circuit consists of one factor of $K$, one factor of $\text{exp}(-ith)$ and one factor of $K^\dagger$, as given in Fig. \EKK{2}(b). Using the Cartan subalgebra given in (\EKK{11}), we see that $\text{exp}(-ith)$ does not require any CNOT gates. Hence, the complete time-evolution circuit of $U(t)=K\text{exp}(-ith)K^\dagger$ has twice as many CNOTs as the circuit for one $K$ has. Therefore, the non-optimized circuit for time evolution has $2n(n^2-1)/3$ CNOTs, whereas the optimized one has only $2n(n-1)$ CNOTs.

\end{document}